\documentclass{article}

\usepackage{arxiv}

\usepackage[utf8]{inputenc} 
\usepackage[T1]{fontenc}    
\usepackage{hyperref}       
\usepackage{url}            
\usepackage{booktabs}       
\usepackage{amsfonts}       
\usepackage{nicefrac}       
\usepackage{microtype}      
\usepackage{lipsum}
\usepackage{appendix}
\usepackage{graphicx}
\usepackage{amsmath}
\usepackage{natbib}
\usepackage{physics}
\usepackage{amssymb}
\usepackage{multirow}
\usepackage[table]{xcolor}
\hypersetup{
  colorlinks,
  citecolor=blue,
  linkcolor=Red,
  urlcolor=Blue}
\usepackage[dvipsnames]{xcolor}
\usepackage{caption}
\usepackage{algorithm}
\usepackage{algpseudocode}
\usepackage{listings}
\usepackage{booktabs}
\usepackage{lipsum}
\usepackage{threeparttable} 

\usepackage[labelfont=bf,labelsep=quad]{caption}  

\graphicspath{ {./images/} }

\title{ \textbf{{ \normalfont  \textbf{reg}}TPS-KLE}: A novel approach to approximate a Gaussian random field for Bayesian spatial modeling}

\author{
 Joaquin Cavieres \\
  Chair of Spatial Data Science and Statistical Learning\\
  Universit\"at G\"ottingen\\
  G\"ottingen, Germany \\
  \texttt{joaquin.caviere@uni-goetingen.de} \\
   \And
  Sebastian Krumscheid\\
  Uncertainty Quantification\\
  Karlsruhe Institute of Technology (KIT)\\
  Karlsruhe, Germany\\
  \texttt{sebastian.krumscheid@kit.edu} \\
}

\begin{document}
\maketitle
\begin{abstract}

Gaussian random field is a ubiquitous model for spatial phenomena in diverse scientific disciplines. Its approximation is often crucial for computational feasibility in simulation, inference, and uncertainty quantification. The Karhunen–Loève Expansion provides a theoretically optimal basis for representing a Gaussian random field as a sum of deterministic orthonormal functions weighted by uncorrelated random variables. While this is a well-established method for dimension reduction and approximation of (spatial) stochastic processes, its practical application depends on the explicit or implicit definition of the covariance structure. In this work, we propose a novel approach, referred to as regTPS-KLE, for approximating a Gaussian random field by explicitly constructing its covariance via a regularized thin plate spline (TPS) kernel. Because TPS kernels are conditionally positive definite and lack a direct spectral decomposition, we formulate the covariance as the inverse of a regularized elliptic operator. To evaluate its statistical performance, we compare its predictive accuracy and computational efficiency with a Gaussian random field approximation constructed using the stochastic partial differential equations (SPDE) method and implemented within an MCMC algorithm.  In simulation studies, the predictive differences between the SPDE and regTPS-KLE models were minimal when the spatial field was generated using Mat\'ern and exponential covariance functions, while regTPS-KLE models consistently outperformed the SPDE approach in terms of computational efficiency. In a real data application, regTPS-KLE exhibits superior predictive accuracy compared with SPDE models based on leave-one-out cross-validation while also achieving improved computational efficiency.

\end{abstract}

\keywords{Gaussian random field \and thin plates splines \and Bayesian inference \and spatial modeling}

\section{Introduction}

Spatial models have attracted growing interest within the scientific community in recent years due to their ability to model spatial dependencies between observations (measurements) in a two-dimensional space, an aspect often inadequately addressed by conventional statistical approaches. In particular, referenced spatial data (also called geostatistical data) involve the analysis of data collected by sampling a continuous spatial process, denoted $u(\mathbf{s}) : \mathbf{s} \in \mathbb{R}^{2}$, at a finite number of spatial locations $\mathbf{s}_i \in \mathcal{D} \subset \mathbb{R}^{2}$, where $\mathbf{s} = (\mathbf{s}_1, \ldots, \mathbf{s}_n)^{\top}$ typically represents coordinates such as latitude and longitude. Because direct observations of $u(\mathbf{s})$ are often subject to measurement error, a common assumption is that these errors are additive. Denoting by $y_i$ the observed values at spatial location $\mathbf{s}_i$, a general spatial model (geostatistical model) can be expressed as:

\begin{equation}\label{eq1}
y_i = \mu + u(\mathbf{s}_i) + \varepsilon_i, \quad i = 1, \ldots, n,
\end{equation}

where the error terms $\varepsilon_i$ are commonly assumed to be independent and identically distributed with zero mean and standard deviation $\sigma_\varepsilon$ \citep{cressie1989geostatistics, moller2013spatial}. This model is easily extended to a regression framework by defining the mean structure as $\mu = \mathbf{x}_i^{\top} \boldsymbol{\beta}$, where $\mathbf{x}_i$ includes explanatory variables related to $y_i$. The primary goal of spatial modeling is to predict the value of the underlying spatial process $u(\mathbf{s})$ at unsampled locations or to estimate the average of $u(\mathbf{s})$ over a region of interest $\mathcal{D}$. A common strategy for handling uncertainty in the continuous spatial stochastic process $u(\mathbf{s})$ is to model it as a Gaussian random field (GRF). Despite its flexibility, assuming a GRF as the spatial random field becomes computationally intensive when dealing with large datasets, motivating the development of scalable approximations. A comprehensive review of scalable spatial models is provided by \citet{heaton2019case}. Notable approaches include low-rank approximations, such as Gaussian predictive process models \citep{banerjee2008gaussian}, as well as methods based on sparse matrix algebra, including the Vecchia approximation \citep{vecchia1988estimation, katzfuss2021general}. One of the most popular approaches to approximating a GRF was proposed by \citet{lindgren2011explicit}, who developed an efficient framework based on stochastic partial differential equations (SPDEs). This approach establishes an explicit link between GRFs and a Gaussian Markov random field (GMRF) \citep{rue2005gaussian}, yielding sparse precision matrices and enabling fast computation without sacrificing accuracy in latent Gaussian models. The SPDE method has been widely adopted across a range of disciplines \citep{lindgren2022spde}. An alternative to GRF-based models is the use of radial basis functions (RBFs; \citealp{wendland2004scattered, wendland2006computational}), and in particular thin plate splines (TPS). Originally introduced by \citet{duchon1977splines}, TPS are spline-based techniques for spatial interpolation and smoothing \citep{wahba1990spline, green1987penalized}. However, TPS suffer from poor scalability, with computational demands increasing rapidly with the number of observations. To overcome this limitation, \citet{wood2003thin} proposed thin plate regression splines (TPRS), which approximate the spline basis using truncated spectral decomposition, yielding improved computational efficiency \citep{wood2017generalized, cavieres2023thin}. Within a Bayesian framework, early work by \citet{wahba1983bayesian} and \citet{nychka1988bayesian} introduced methods to construct credible intervals for smoothing splines. Building on these ideas and the correspondence between spline smoothing and stochastic processes established by \citet{kimeldorf1970correspondence}, \citet{white2006bayesian} derived a spatial prior based on TPS, enabling Bayesian inference without resorting to Markov chain Monte Carlo methods. For broader perspectives on Bayesian spatial modeling, see \citet{handcock1993bayesian}, \citet{gelfand2012hierarchical}, and \citet{banerjee2014hierarchical}. In spatial statistics and uncertainty quantification, the Karhunen--Lo\`eve expansion (KLE) is a widely used spectral method for approximating GRFs \citep{lord2014introduction, alexanderian2015brief}. The KLE provides a truncated series representation of a GRF in terms of eigenfunctions and eigenvalues derived from the covariance kernel. Early theoretical foundations were laid by \citet{loeve1978probability} and \citet{adler1981geometry}, with later applications in stochastic finite element methods \citep{ghanem1991stochastic}. In geostatistics, KLE has been used to model spatial variability through orthogonal decompositions of covariance structures \citep{zhu1997karhunen, brenner2017karhunen}. Further applications include environmental modeling, Bayesian inverse problems \citep{marzouk2009stochastic, biegler2011large}, Gaussian process emulation \citep{higdon2008computer}, and hierarchical Bayesian frameworks \citep{cotter2010approximation}. Despite its advantages, the KLE requires the solution of an often expensive eigenvalue problem, motivating ongoing research into efficient numerical approximations, including adaptive and sparse variants \citep{le2010spectral, litvinenko2022karhunen}.

In this work, we propose a novel methodology to approximate a GRF based on the KLE, where the covariance function (covariance kernel) is defined through the inverse of a regularized elliptic operator. We refer to this approach as regTPS-KLE. The proposed framework establishes a direct and explicit connection between classical spline smoothing theory and spatial stochastic process modeling, yielding a covariance-driven KLE whose modes inherit the analytical and smoothing properties of thin plate splines. The main contributions of this work can be summarized as follows:

\begin{enumerate}
\item We develop a theoretical approximation of a GRF covariance function defined as the inverse of a regularized elliptic operator,

\begin{equation*}
L_{\alpha} = \mathbf{I} + \alpha \Delta^2.    
\end{equation*}

While the Bayesian interpretation of penalized regression is well established, our approach explicitly derives the associated spectral representation using Hilbert-Schmidt theory and Mercer's theorem. This provides a rigorous foundation for the KLE and clarifies its relationship with the TPS geometry. The resulting probabilistic formulation unifies penalized regression and spatial process modeling, enabling principled uncertainty quantification through the full Bayesian posterior over latent coefficients and hyperparameters.

\item Through regularization, we convert the conditionally positive definite TPS kernel into a strictly positive definite Hilbert-Schmidt kernel, thereby ensuring the validity of the Mercer decomposition. In contrast to standard constructions, we provide an explicit spectral treatment of the polynomial null space: eigenfunctions with zero bending energy are retained and assigned unit prior variance, rather than being implicitly constrained or projected out.

\item The KLE basis functions are obtained from the eigensystem of the regularized elliptic operator and inherit the smoothness hierarchy induced by the bending-energy functional. The resulting eigenvalue ordering naturally separates unpenalized polynomial trends from increasingly penalized high-frequency modes, providing both theoretical justification and practical guidance for adaptive truncation of the expansion.

\item The proposed regTPS-KLE framework produces spatially smooth realizations by construction, as smoothness is directly encoded in the operator-based prior via the bending-energy penalty. This contrasts with many parametric covariance models, such as the Mat\'ern class, where smoothness must be carefully controlled through additional parameters, and provides a natural prior for applications requiring guaranteed regularity of spatial processes.

\item By explicitly assigning unit prior variance to the polynomial null space, the regTPS-KLE approach accommodates large-scale spatial trends without imposing smoothing penalties. This treatment constitutes a theoretical and practical advantage over standard KLE formulations, in which the null space is often discarded or handled inadequately.
\end{enumerate}

For comparative purposes, we conduct spatial modeling using the widely adopted SPDE approach by Markov chain Monte Carlo (MCMC) sampling, considering both simulated data and a real data application. The remainder of the paper is organized as follows. Section~\ref{section2} reviews GRFs and KLE theories, followed by a detailed discussion of TPS and their connection to Hilbert-Schmidt kernels, where the proposed regTPS-KLE methodology is introduced. In Section~\ref{section3}, these theoretical developments are extended to a Bayesian framework for spatial modeling. Section~\ref{section4} presents numerical analyses on simulated data, including comparisons with the SPDE approach in terms of root mean squared error (RMSE), mean absolute error (MAE), and $R^2$, as well as computational efficiency, under various scenarios and covariance specifications. In Section~\ref{section5}, the SPDE and regTPS-KLE approaches are applied to NO$_2$ concentration data in Germany, with predictive performance evaluated using leave-one-out cross-validation and computational efficiency assessed. In Section~\ref{section6}, we discuss the results of our analyses, and Section~\ref{section7} concludes the paper.

\section{Methodology}
\label{section2}

\subsection{Gaussian Random Field}
\label{subsection2.1}

Let $(\Omega, \mathcal{F}, \mathbf{P})$ be a probability space, and let $\mathcal{D} \subseteq \mathbb{R}^d$ be a bounded index set representing the spatial domain, commonly $d=2$ for the spatial domain. A spatial random field can be defined as a function $\mathbf{u}(\mathbf{s}, \omega): \mathcal{D} \times \Omega \rightarrow \mathbb{R}$, where $\mathbf{s} \in \mathcal{D}$ is a vector of spatial coordinates $\mathbf{s} = (s_1, s_2)$, and $\omega \in \Omega$ a generic outcome of the sample space. Furthermore, a spatial random field is a collection of random variables at each spatial coordinate $\mathbf{s}$ in $\mathcal{D}$. Now, let's consider a set of spatial locations $\mathbf{s} = (\mathbf{s}_1, \ldots, \mathbf{s}_N)^{\top}$ associated with a set of random variables $\mathbf{u}(\mathbf{s}, \omega) = \{u(\mathbf{s}_1, \omega), \ldots, u(\mathbf{s}_{N}, \omega)\}$, where $N$ is the number of observed random variables. For simplicity in notation, we only will consider to $\mathbf{u}(\mathbf{s})$ as the spatial random field instead of $\mathbf{u}(\mathbf{s}, \omega)$. A spatial random field is called a \textit{Gaussian} random field (GRF) when any finite set of its realizations $\{u(\boldsymbol{s}_{1}), \ldots, u(\boldsymbol{s}_{N})\}$ follows a multivariate normal distribution, i.e., $\{u(\boldsymbol{s}_{1}), \ldots, u(\boldsymbol{s}_{N})\} \sim \mathcal{N}_{N}(\boldsymbol{\mu}, \Sigma)$ \citep{adler2010geometry,lord2014introduction}. A GRF is entirely defined by its mean function $\mu(\boldsymbol{s}) = \mathbb{E}[u(\boldsymbol{s})]$ and its covariance function $C(\mathbf{s}, \mathbf{s}^{'}) = \text{Cov}(u(\mathbf{s}), u(\mathbf{s}^{'}))$. Moreover, a GRF is stationary if the associated finite-dimensional distributions of the field are invariant under shifts in space, and isotropic, if the covariance between any two points depends solely on the distance separating them, meaning $C(\mathbf{s}, \mathbf{s}^{'}) = C(\mathbf{d})$, where $\mathbf{d} = \|\mathbf{s} - \mathbf{s}^{'}\|$ (Euclidean distance) \citep{abrahamsen1997review}.

\subsection{The Karhunen-Loève Expansion}
\label{subsection2.2}

Consider a second-order (finite variance) spatial random field $u(\mathbf{s})$ with mean $\mu(\mathbf{s})$ and covariance $C(\mathbf{s}, \mathbf{s}')$. Covariance functions are symmetric and positive semidefinite functions that belong to the class of Hilbert-Schmidt kernels (in the Euclidean space, these are functions with finite Frobenius norm; \citet{abrahamsen1997review}). These properties ensure that the associated covariance operator admits a countable orthonormal basis of eigenfunctions, with the corresponding eigenvalues being real and non-negative \citep{uribe2020bayesian}. Considering Mercer's theorem \citep{mercer1909xvi}, the covariance kernel can be represented by a series expansion based on the spectral representation of the covariance operator \citep{kolmogorov1975introductory} as:

\begin{equation}\label{eq2}
C(\mathbf{s}, \mathbf{s}') = \sum^{\infty}_{k=1}\lambda_k\phi_k(\mathbf{s})\phi_k(\mathbf{s}'),
\end{equation}

where $\lambda_k \in [0, \infty)$ and $\phi_k(\mathbf{s}): \mathcal{D} \rightarrow \mathbb{R}$ (where $\phi_{k}(\mathbf{s}) \in L^{2}(\mathcal{D})$, with $L^{2}(\mathcal{D})$ as the Hilbert space of square-integrable functions \citep{uribe2020bayesian}). In this way, a spatial random field $u(\mathbf{s})$ can be approximated by $\tilde{u}(\mathbf{s})$ using the KLE as an infinite series of orthogonal functions (eigenfunctions) and uncorrelated random variables as:

\begin{equation}\label{eq3}
u(\mathbf{s}) \approx \tilde{u}(\mathbf{s}) = \mu(\mathbf{s}) + \sum_{k=1}^{M} \sqrt{\lambda_k} \, \phi_k(\mathbf{s}) \, \epsilon_k,
\end{equation}

where $u(\mathbf{s})$ is the GRF at spatial location $\mathbf{s}$, $\mu(\mathbf{s})$ is the mean function of the GRF, $\lambda_k$ are the eigenvalues, $\phi_k(\mathbf{s})$ are the eigenfunctions forming an orthonormal basis, and $\epsilon_k$ are uncorrelated i.i.d.\ random errors with $\epsilon_k \sim \mathcal{N}(0, 1)$ \citep{fasshauer2015kernel, lord2014introduction}. The eigenfunctions $\phi_k(\mathbf{s})$ and the eigenvalues $\lambda_k$ are determined by the covariance function $C(\mathbf{s}, \mathbf{s}')$ of the GRF, specifically by solving the following Fredholm integral eigenvalue problem over the spatial domain $\mathcal{D}$:

\begin{equation}\label{eq4}
\int_{\mathcal{D}} C(\mathbf{s}, \mathbf{s}') \phi_k(\mathbf{s}') \, d\mathbf{s}' = \lambda_k \phi_k(\mathbf{s}),
\end{equation}

whose analytical solution exists only for specific cases of covariance functions \citep{ghanem2003stochastic}. Covariance kernels play a central role in random field modeling by characterizing the spatial correlation structure of the field. Among the widely used and flexible families of isotropic covariance functions are the Whittle--Mat\'ern kernels \citep{matern1960spatial}, which form a class of Hilbert-Schmidt kernels. These kernels provide a tunable framework to model varying degrees of smoothness and spatial dependence, and are mathematically defined as

\begin{equation}\label{eq5}
C(\mathbf{d}) = \sigma^{2}_{u}\frac{2^{1-\nu}}{\Gamma(\nu)}\left(\frac{\sqrt{2\nu\mathbf{d}}}{\rho}\right)^{\nu}K_{\nu}\left(\frac{\sqrt{2\nu\mathbf{d}}}{\rho}\right),
\end{equation}

where $\Gamma(\cdot)$ is the gamma function, $K_{\nu}(\cdot)$ is the modified Bessel function of the second kind, $\rho$ is the range parameter (correlation length), and $\nu > 0$ is the smoothness parameter. Particular cases of Whittle--Mat\'ern kernels include $C_{1/2}(\mathbf{d}) = \sigma^{2}_{u} \exp\left(-\frac{\mathbf{d}}{\rho}\right)$, called the \textit{exponential} kernel, and $C_{\infty}(\mathbf{d}) = \sigma^{2}_{u} \exp\left(-\frac{\mathbf{d}^{2}}{2\rho^{2}}\right)$, called the \textit{squared exponential} (also known as Gaussian) covariance kernel, respectively \citep{banerjee2014hierarchical, van2019theory}. 

\subsection{Thin Plate Splines as a Hilbert-Schmidt Kernel}

Consider a set of spatial data points $\{(\mathbf{s}_i, y_i)\}_{i=1}^N$, where $\mathbf{s}_i = (s_{i1}, s_{i2}) \in \mathcal{D} \subset \mathbb{R}^{2}$ represent the spatial locations and $y_i \in \mathbb{R}$ are the observed values. Our main objective is to construct an interpolant $f: \mathcal{D} \rightarrow \mathbb{R}$ such that $f(\mathbf{s}_i) = y_i$ for $i = 1, \ldots, N$. This interpolant is built as a linear combination of \textit{radial basis functions} (RBFs) centered at the spatial data points such that:

\begin{equation}\label{eq6}
f(\mathbf{s}) = \sum_{i=1}^{N} c_i \phi(\norm{\mathbf{s} - \mathbf{s}_i}),
\end{equation}

where $\phi: [0, \infty) \rightarrow \mathbb{R}$ is the RBF, $\norm{\cdot}$ denotes the Euclidean norm, $c_i$ are the coefficients to be determined, and $\mathbf{s}_i \in \mathbb{R}^d$ are the known spatial locations. Originally introduced by \citet{duchon1977splines}, thin plate splines (TPS) is a special case of RBF interpolation derived from variational principles and widely used for smooth surface fitting. For TPS, the RBF (aka kernel function) is $\phi(\mathbf{d}) = \mathbf{d}^{2} \log(\mathbf{d})$, where the natural setting for a smoothing TPS is the Sobolev space $H^{m}(\mathcal{D})$. For a smoothing TPS defined in a two-dimensional space, the bending energy is defined for functions $f \in H^{2}(\mathcal{D})$. In this way, for $L^{2}(\mathcal{D})$ (the space of square-integrable functions with inner product $\langle f, g\rangle_{L^{2}}= \int_{\mathcal{D}}f(\mathbf{s})g(\mathbf{s})\,d\mathbf{s}$) and $H^{2}(\mathcal{D})$ (the space of functions in $L^{2}(\mathcal{D})$), the bending energy seminorm $J(f)$ for a two-dimensional TPS is given by \citep{wahba1990spline, green1993nonparametric, cavieres2023thin}:
\begin{equation}\label{eq7}
J(f) = \int_{\mathcal{D}} \left[ \left(\frac{\partial^{2} f}{\partial s_{1}^{2}}\right)^{2} + 2 \left(\frac{\partial^{2} f}{\partial s_{1} \partial s_{2}}\right)^{2} + \left(\frac{\partial^{2} f}{\partial s_{2}^{2}}\right)^{2} \right] d\mathbf{s} = \int_{\mathcal{D}} |\Delta f(\mathbf{s})|^{2}\, d\mathbf{s},
\end{equation}
where $\Delta$ is the Laplacian operator and $J(f)$ is a seminorm on $H^{2}(\mathcal{D})$. Specifically, $J(f) = 0$ if and only if $f(\mathbf{s})$ is a polynomial of total degree less than 2 (i.e., $f(\mathbf{s}) = c_0 + c_1 s_1 + c_2 s_2$) \citep{berlinet2011reproducing, fasshauer2015kernel}. Denoting this polynomial space as $\mathcal{P}_1$, then $\mathcal{P}_1$ is the null space of the bending energy operator. Finally, the TPS aims to find $f \in H^{2}(\mathcal{D})$ minimizing the following functional:
\begin{equation}\label{eq8}
\mathcal{L}(f) = \sum_{i=1}^{N} (y_i - f(\mathbf{s}_i))^2 + \alpha J(f),
\end{equation}
where $\alpha > 0$ is a smoothing parameter and $J(f)$ was defined in Equation \eqref{eq7}.

\newpage
\subsubsection{Thin Plate Splines as a Green's Function Solution}

A Green's function is a fundamental solution to a linear differential operator. Essentially, it allows us to solve linear partial differential equations (PDEs) by converting them into integral equations \citep{hilbert1985methods, wahba1990spline, fasshauer2015kernel}. The formal definition is the following: given a linear differential operator $\mathcal{L}$ on the domain $\mathcal{D} \subset \mathbb{R}^{d}$, the Green's kernel $G$ of $\mathcal{L}$ is defined as the solution of:

\begin{equation}\label{eq9}
 \mathcal{L}G(\mathbf{s}, \mathbf{s}') = \delta(\mathbf{s} - \mathbf{s}'), \hspace{3mm} \mathbf{s}' \in \mathcal{D} \hspace{1mm} \text{fixed}.
\end{equation}

Here $\delta(\mathbf{s} - \mathbf{s}')$ is the Dirac delta functional evaluated at $\mathbf{s} - \mathbf{s}'$. In this case, $\delta$ acts as a point evaluator for any $f \in L^{2}(\mathcal{D})$, i.e., 

\begin{equation}\label{eq10}
\int_{\mathcal{D}} f(\mathbf{s}')\delta(\mathbf{s} - \mathbf{s}')d\mathbf{s}' = f(\mathbf{s})
\end{equation}

\citep{duffy2015green, evans2022partial}. Note that, although this property of the Dirac delta function is analogous to the reproducing property of a reproducing kernel $K$ \citep{fasshauer2015kernel}, $\delta$ is not the reproducing kernel of $L^{2}(\mathcal{D})$ since $\delta \notin L^{2}(\mathcal{D})$. For the biharmonic operator $\Delta^{2}$, the Green's function $G(\mathbf{s}, \mathbf{s}')$ satisfies:

\begin{equation}\label{eq11}
\Delta^{2}G(\mathbf{s}, \mathbf{s}') = \delta(\mathbf{s} - \mathbf{s}') , \quad \mathbf{s}, \mathbf{s}' \in \mathbb{R}^{d}.
\end{equation}

For $\mathbb{R}^{2}$, the biharmonic Green's function is given by \citep{aronszajn1957characterization, duffy2015green, fasshauer2015kernel}:

\begin{equation}\label{eq12}
G(\mathbf{s}, \mathbf{s}') = \frac{1}{8\pi}\norm{\mathbf{s} - \mathbf{s}'}^{2}\log \norm{\mathbf{s} - \mathbf{s}'}.
\end{equation}

Since the scale of the kernel is absorbed into $\alpha$ and the coefficients $c_i$, the TPS kernel $K(\mathbf{s}, \mathbf{s}') = \norm{\mathbf{s} - \mathbf{s}'}^{2} \log \norm{\mathbf{s} - \mathbf{s}'}$ is the fundamental solution (Green's function) for the biharmonic operator on $\mathbb{R}^{2}$. 

\subsubsection{Hilbert-Schmidt Operators and Kernels}

A linear operator $T: L^{2}(\mathcal{D}) \rightarrow L^{2}(\mathcal{D})$ is a Hilbert-Schmidt operator if its Hilbert-Schmidt norm is finite:

\begin{equation}\label{eq13}
\norm{T}^{2}_{\text{HS}} = \sum^{\infty}_{k=1}\norm{Te_{k}}^{2}_{L^{2}} < \infty,
\end{equation}

for any orthonormal basis $\{e_k\}$ of $L^{2}(\mathcal{D})$ \citep{reed1980methods, engl2015regularization}. If $T$ is an integral operator with kernel $K(\mathbf{s}, \mathbf{s}')$:

\begin{equation}
(Tf)(\mathbf{s}) = \int_{\mathcal{D}} K(\mathbf{s}, \mathbf{s}') f(\mathbf{s}') d\mathbf{s}',
\end{equation}

then $T$ is a Hilbert-Schmidt operator if and only if the kernel $K(\mathbf{s}, \mathbf{s}')$ is square-integrable, i.e., $K \in L^2(\mathcal{D} \times \mathcal{D})$.

To recap, the main requirement for the KLE is a covariance function $C(\mathbf{s}, \mathbf{s}')$ that is symmetric and positive semi-definite. However, the TPS kernel $K(\mathbf{s}, \mathbf{s}') = \mathbf{d}^{2} \log(\mathbf{d})$ is only conditionally positive definite \citep{fasshauer2007meshfree, wendland2004scattered}. To utilize the spectral properties required for Bayesian inference, we must reframe the problem. Instead of working with the singular biharmonic operator directly, we define a regularized operator $L_{\alpha} = \mathbf{I} + \alpha \Delta^2$ that yields a proper Hilbert-Schmidt kernel.

\subsubsection{Connecting Thin Plate Splines to Hilbert-Schmidt Kernels}

To obtain a valid Hilbert-Schmidt kernel associated with TPS, we introduce a regularized elliptic operator $L_{\alpha} = \mathbf{I} + \alpha \Delta^{2}$, where $\alpha > 0$ is a regularization parameter and $\mathbf{I}$ denotes the identity operator on the Sobolev space $H^{2}(\mathcal{D})$. The operator $L_{\alpha}$ is elliptic, self-adjoint, and invertible. Its inverse, $L_{\alpha}^{-1} = (\mathbf{I} + \alpha \Delta^{2})^{-1}$, defines a compact, self-adjoint, and strictly positive definite integral operator. In the context of continuous variational problems, Mercer's theorem \citep{mercer1909xvi, steinwart2012mercer} and Hilbert-Schmidt theory \citep{renardy2004introduction} apply to compact, self-adjoint operators with positive definite kernels. As we already know, TPS is conditionally positive definite and therefore fall outside the direct scope of these results. To address this issue, we reformulate the problem by replacing the TPS kernel $K(\mathbf{s}, \mathbf{s}')$ with the kernel induced by the inverse of the regularized operator $L_{\alpha}$. Let's consider the following continuous variational problem:

\begin{equation}\label{eq14}
\underset{f \in H^{2}(\mathcal{D})}{\text{min}}\left(\int_{\mathcal{D}} (f(\mathbf{s}) - g(\mathbf{s}))^2 \, d\mathbf{s} + \alpha J(f)\right),
\end{equation}

where $g(\mathbf{s})$ is a target function, $\mathcal{D} \subset \mathbb{R}^d$ is a bounded domain, and $J(f) = \int_{\mathcal{D}} (\Delta^2 f)^2 \, d\mathbf{s}$ is the bending-energy functional. Applying the Euler--Lagrange equation to this continuous problem yields the necessary condition for a minimizer:

\begin{equation}\label{eq15}
f(\mathbf{s}) - g(\mathbf{s}) + \alpha \Delta^{2}f(\mathbf{s}) = 0.
\end{equation}

Rearranging, we obtain the elliptic partial differential equation (PDE):

\begin{align}\label{eq16}
f(\mathbf{s}) + \alpha \Delta^{2}f(\mathbf{s}) &= g(\mathbf{s}), \\ \nonumber
L_{\alpha}f &= g,
\end{align}

which is a fourth-order elliptic PDE \citep{trudinger1983elliptic, courant2024methods}. Under appropriate boundary conditions—such as clamped plate conditions (fixed values and normal derivatives on $\partial\mathcal{D}$)—the biharmonic operator $\Delta^{2}$ is an unbounded, self-adjoint operator on $L^2(\mathcal{D})$ with domain $H^4(\mathcal{D}) \cap H^2_0(\mathcal{D})$. The regularized operator $L_\alpha$ is then boundedly invertible, and its inverse $L_\alpha^{-1}$ is a compact operator. Most importantly, the integral kernel associated with $L_{\alpha}^{-1}$, denoted $K_{\alpha}(\mathbf{s}, \mathbf{s}')$, is a well-defined Hilbert-Schmidt kernel satisfying:

\begin{equation}\label{eq17}
(L_\alpha^{-1} g)(\mathbf{s}) = \int_{\mathcal{D}} K_\alpha(\mathbf{s}, \mathbf{s}') g(\mathbf{s}') \, d\mathbf{s}'.
\end{equation}

Furthermore, by Mercer's theorem, the kernel $K_\alpha$ admits the spectral expansion:

\begin{equation}\label{eq18}
K_\alpha(\mathbf{s}, \mathbf{s}') = \sum_{k=1}^{\infty} \lambda_{k,\alpha} \, \phi_k(\mathbf{s}) \, \phi_k(\mathbf{s}'),
\end{equation}

where $\{\lambda_{k,\alpha}\}_{k=1}^\infty$ are the eigenvalues and $\{\phi_k(\mathbf{s})\}_{k=1}^\infty$ are the corresponding orthonormal eigenfunctions of the compact operator $L^{-1}_{\alpha}$, satisfying:

\begin{equation}\label{eq19}
\int_{\mathcal{D}} \phi_k(\mathbf{s}) \phi_\ell(\mathbf{s}) \, d\mathbf{s} = \delta_{k\ell},
\end{equation}

where $\delta_{k\ell}$ is the Kronecker delta, and with $\sum_{k=1}^\infty \lambda_{k,\alpha}^2 < \infty$ (Hilbert-Schmidt property). These eigenfunctions are solutions to:

\begin{equation}\label{eq20}
L_\alpha^{-1} \phi_k(\mathbf{s}) = \lambda_{k,\alpha} \, \phi_k(\mathbf{s}),
\end{equation}

or equivalently, applying $L_\alpha$ to both sides, we obtain the eigenvalue problem for the operator $L_\alpha$:

\begin{equation}\label{eq21}
(\mathbf{I} + \alpha \Delta^{2}) \phi_k(\mathbf{s}) = \mu_k(\alpha) \, \phi_k(\mathbf{s}),
\end{equation}

where $\mu_k(\alpha) = 1/\lambda_{k,\alpha}$ denotes the eigenvalues of the operator $L_\alpha$ itself (not its inverse). Let $\{v_k\}_{k=1}^\infty$ denote the eigenvalues of the biharmonic operator $\Delta^{2}$, ordered in ascending order, so that $\Delta^{2} \phi_k(\mathbf{s}) = v_k \phi_k(\mathbf{s})$. Substituting this into Equation~\eqref{eq21}, we obtain:

\begin{equation}\label{eq22}
\phi_k(\mathbf{s}) + \alpha v_k \phi_k(\mathbf{s}) = \mu_k(\alpha) \, \phi_k(\mathbf{s}),
\end{equation}

which simplifies to:

\begin{equation}\label{eq23}
(1 + \alpha v_k) \phi_k(\mathbf{s}) = \mu_k(\alpha) \, \phi_k(\mathbf{s}).
\end{equation}

This establishes the relationship between the eigenvalues:

\begin{equation}\label{eq24}
\mu_k(\alpha) = 1 + \alpha v_k,
\qquad \text{and consequently} \qquad
\lambda_{k,\alpha} = \frac{1}{1 + \alpha v_k}.
\end{equation}

The eigenvalue spectrum of $\Delta^{2}$ has a special structure. The biharmonic operator $\Delta^2$ has a null space $\mathcal{N}(\Delta^2)$ consisting of polynomial functions of degree at most $m-1$ (where $m=2$ for standard TPS, yielding polynomials of degree up to 1). In two dimensions, this null space has dimension $d_{\text{null}} = 3$, spanned by the constant function and the linear terms $x$ and $y$. For these polynomial functions, we have $\Delta^2 p(\mathbf{s}) = 0$, which corresponds to $v_k = 0$ for $k \leq d_{\text{null}}$. Thus, the eigenvalue structure can be characterized as follows:

\begin{itemize}
    \item \textbf{Null space modes} ($k \leq d_{\text{null}}$, $v_k = 0$): These eigenvalues correspond to the polynomial null space $\mathcal{N}(\Delta^2)$. For these modes, 
    \begin{equation*}
    \mu_k(\alpha) = 1 + \alpha \cdot 0 = 1,
    \qquad
    \lambda_{k,\alpha} = \frac{1}{1} = 1.
    \end{equation*}
    This implies that $L^{-1}_{\alpha}$ maps these polynomials to themselves with unit eigenvalue, and they are \emph{unpenalized} by the bending-energy functional. The eigenfunctions $\phi_k$ for $k \leq d_{\text{null}}$ span the polynomial space $\mathcal{P}_{m-1}(\mathcal{D})$ and capture large-scale spatial trends with unit prior variance.
    
    \item \textbf{Penalized modes} ($k > d_{\text{null}}$, $v_k > 0$): These eigenvalues correspond to the non-polynomial bending modes of the plate, which are orthogonal to the null space in the $L^2$ inner product. For these modes,
    \begin{equation*}
    \mu_k(\alpha) = 1 + \alpha v_k > 1,
    \qquad
    \lambda_{k,\alpha} = \frac{1}{1 + \alpha v_k} < 1.
    \end{equation*}
    The eigenvalues $v_k$ increase with $k$, corresponding to increasingly oscillatory eigenfunctions with higher bending energy. The regularization parameter $\alpha$ controls the decay rate of $\lambda_{k,\alpha}$, where larger $\alpha$ causes faster decay, imposing stronger smoothing on high-frequency components.
\end{itemize}

In the classical TPS solution framework, the polynomial part is handled separately from the radial basis function part. The coefficients are typically constrained such that the radial basis contribution is orthogonal to the null space of polynomials, ensuring well-posedness and uniqueness of the ``bending'' part of the solution \citep{wahba1975smoothing, green1987penalized, cavieres2023thin}. The regularization parameter $\alpha > 0$ makes the entire problem well-posed without requiring explicit orthogonality constraints, as the operator $L_\alpha = \mathbf{I} + \alpha \Delta^2$ is boundedly invertible. Therefore, the kernel of the regularized TPS, denoted $K_{\alpha}(\mathbf{s}, \mathbf{s}')$ and associated with the inverse operator $L^{-1}_{\alpha} = (\mathbf{I} + \alpha \Delta^{2})^{-1}$, is a Hilbert-Schmidt, self-adjoint, positive definite kernel with spectral expansion:

\begin{equation}\label{eq25}
K_{\alpha}(\mathbf{s}, \mathbf{s}') = \sum_{k=1}^{\infty}\lambda_{k,\alpha} \, \phi_k(\mathbf{s}) \, \phi_k(\mathbf{s}'),
\end{equation}

where $\lambda_{k,\alpha} = (1 + \alpha v_k)^{-1}$ and $\{\phi_k(\mathbf{s})\}_{k=1}^\infty$ are the complete orthonormal system of eigenfunctions of $\Delta^{2}$ (equivalently, of $L_\alpha$ and $L_\alpha^{-1}$), including both the polynomial null space modes and the orthogonal complement. This spectral representation forms the theoretical foundation of the regTPS-KLE approach for practical Bayesian inference on spatial data described in Section~\ref{section3}.

\section{Approximating a GRF by a regTPS-KLE: A Bayesian Approach}
\label{section3}

In the previous section, we established that the regularized TPS covariance kernel admits the spectral expansion:

\begin{equation}\label{eq27}
K_{\alpha}(\mathbf{s}, \mathbf{s}') = \sum_{k=1}^{\infty} \lambda_{k,\alpha} \, \phi_k(\mathbf{s}) \, \phi_k(\mathbf{s}'),
\end{equation}

where $\phi_k(\mathbf{s})$ are the eigenfunctions of the regularized biharmonic operator
$L_{\alpha} = \mathbf{I} + \alpha\Delta^2$, satisfying the eigenvalue problem:

\begin{equation}\label{eq28}
(\mathbf{I} + \alpha\Delta^2)\phi_k(\mathbf{s}) = \mu_k(\alpha) \, \phi_k(\mathbf{s}),
\end{equation}

where $\mu_k(\alpha) = 1 + \alpha v_k$ are the eigenvalues of the operator $L_\alpha$, and $v_k$ are the eigenvalues of the biharmonic operator $\Delta^2$. The corresponding covariance eigenvalues (KLE prior variances) are:

\begin{equation}\label{eq29}
\lambda_{k,\alpha} = \frac{1}{\mu_k(\alpha)} = \frac{1}{1 + \alpha v_k}.
\end{equation}

While this continuous formulation provides a theoretical foundation, the eigenfunctions
$\phi_k$ are generally not available in closed form for arbitrary domains
$\mathcal{D} \subset \mathbb{R}^d$ and boundary conditions. To obtain a computationally tractable framework,
we employ the Ritz--Galerkin approximation by projecting the infinite-dimensional problem onto
a finite-dimensional subspace spanned by TPS basis functions.

\subsection{Finite-Dimensional Approximation via TPS Basis Functions}

Let $N$ denote the number of spatial observations at locations $\{\mathbf{s}_i\}_{i=1}^N \subset \mathcal{D}$, and let $K$ denote the number of TPS basis functions used in the approximation. We introduce a finite-dimensional approximation space $\mathcal{V}_K$ spanned by
$K$ TPS basis functions:

\begin{equation}\label{eq30}
\mathcal{V}_K =
\text{span}\{\tilde{\phi}_1, \tilde{\phi}_2, \ldots, \tilde{\phi}_K\},
\quad \text{where} \quad
\tilde{\phi}_j(\mathbf{s}) = \phi_{\text{TPS}}(\|\mathbf{s} - \boldsymbol{\xi}_j\|),
\end{equation}

and $\phi_{\text{TPS}}(d)$ is the TPS basis function, with knot locations $\{\boldsymbol{\xi}_j\}_{j=1}^K$ typically chosen as a subset of the spatial locations or placed on a regular grid. At this point, it is important to distinguish clearly between two types of functions:

\begin{itemize}
    \item $\phi_k(\mathbf{s})$: the $k$-th eigenfunction of the continuous regularized operator
    $L_{\alpha}$, belonging to the infinite-dimensional Sobolev space
    $H^2(\mathcal{D})$.
    \item $\tilde{\phi}_j(\mathbf{s})$: the $j$-th TPS basis function, which is
    \emph{not} an eigenfunction but serves as a computational basis element for constructing the approximation space $\mathcal{V}_K$.
\end{itemize}

The connection between these functions is established through the Galerkin projection.
Since TPS basis functions form a dense subset of $H^2(\mathcal{D})$
\citep{duchon1977splines}, any eigenfunction $\phi_k \in H^2(\mathcal{D})$ can be
approximated within $\mathcal{V}_K$ by a linear combination:

\begin{equation}\label{eq31}
\phi_k(\mathbf{s}) \approx \hat{\phi}_k(\mathbf{s})
= \sum_{j=1}^{K} a_{kj} \, \tilde{\phi}_j(\mathbf{s}),
\end{equation}

where $\hat{\phi}_k \in \mathcal{V}_K$ denotes the discrete approximation to $\phi_k$.
The approximation error $\|\phi_k - \hat{\phi}_k\|_{H^2(\mathcal{D})} \to 0$ as $K \to \infty$, provided the knot points $\{\boldsymbol{\xi}_j\}$ become dense in $\mathcal{D}$ and the domain satisfies standard smoothness conditions \citep{brenner2008mathematical}.

\subsection{Discrete Eigenvalue Problem}

The coefficients $\mathbf{a}_k = (a_{k1}, \ldots, a_{kK})^{\top}$ are obtained by
diagonalizing the discrete bending-energy operator associated with $\Delta^2$.
This operator is represented in $\mathcal{V}_K$ by the penalty matrix $\mathbf{S} \in \mathbb{R}^{K \times K}$, whose entries are
defined by the bending-energy bilinear form:

\begin{equation}\label{eq32}
\mathbf{S}_{jl}
= \int_{\mathcal{D}}
(\Delta \tilde{\phi}_j(\mathbf{s}))
(\Delta \tilde{\phi}_l(\mathbf{s})) \, d\mathbf{s}.
\end{equation}

This matrix provides a finite-dimensional representation of the biharmonic operator $\Delta^2$ restricted to $\mathcal{V}_K$ and defines the natural geometry for TPS smoothing, the reproducing kernel Hilbert space (RKHS) associated with the bending-energy seminorm \citep{wahba1990spline, brenner2008mathematical}. In practice, the penalty matrix $\mathbf{S}$ is constructed using the method described in \citet{wood2017generalized}, which provides numerically stable computation from the TPS basis representation without requiring explicit evaluation of the integral in Equation~\eqref{eq32}. Here, we perform the standard eigendecomposition of $\mathbf{S}$:

\begin{equation}\label{eq33}
\mathbf{S} = \mathbf{\Psi}_S \mathbf{V}_S \mathbf{\Psi}_S^{\top},
\end{equation}

where $\mathbf{V}_S = \text{diag}(v_1, v_2, \ldots, v_K)$ contains the eigenvalues of $\mathbf{S}$ ordered in ascending order, and the columns of $\mathbf{\Psi}_S \in \mathbb{R}^{K \times K}$ contain the corresponding eigenvectors, which are orthonormal: $\mathbf{\Psi}_S^{\top}\mathbf{\Psi}_S = \mathbf{I}_K$. The eigenvalues $v_k$ of the discrete penalty matrix $\mathbf{S}$ provide finite-dimensional approximations to the eigenvalues of the continuous biharmonic operator $\Delta^2$ restricted to $\mathcal{V}_K$, with the approximation improving as $K$ increases and the basis becomes dense. This standard eigenvalue problem corresponds to an implicit diagonal approximation of the $L^2$ mass matrix, which is computationally efficient and widely used in TPS software implementations \citep{wood2017generalized}. A fully consistent Galerkin discretization would require solving a generalized eigenvalue problem incorporating an explicit mass matrix to preserve $L^2$ orthogonality of the discrete eigenfunctions. We provide a detailed theoretical justification for the standard eigenvalue approximation in Appendix~\ref{secA1}.

\subsubsection{Eigenvalue Ordering and the Polynomial Null Space}

A critical feature of the TPS penalty matrix is its null space, which corresponds to the space of polynomial functions of degree $m-1$ in dimension $d$ (where $m=2$ for standard TPS in two-dimensional space, yielding polynomials up to degree 1). This null space has dimension $d_{\text{null}} = \binom{d+m-1}{d}$ (e.g., $d_{\text{null}} = 3$ for two-dimensional space, corresponding to the constant, $x$, and $y$ polynomial terms). The penalty matrix $\mathbf{S}$ assigns zero bending energy to these polynomial functions, resulting in $d_{\text{null}}$ eigenvalues that are numerically zero (within machine precision $\epsilon \approx 10^{-12}$). We order the eigenvalues in \emph{ascending} order:

\begin{equation}\label{eq34}
v_1 \leq v_2 \leq \cdots \leq v_K,
\end{equation}

such that

\begin{equation}\label{eq35}
v_k < \epsilon \quad \text{for } k \leq d_{\text{null}},
\quad \text{and} \quad
v_k > 0 \quad \text{for } k > d_{\text{null}}.
\end{equation}

The first $d_{\text{null}}$ eigenvectors span the polynomial null space and represent unpenalized large-scale trends in the spatial random field. The remaining eigenvectors correspond to increasingly rough, high-frequency spatial features that are penalized by the bending-energy norm. Specifically, larger eigenvalues $v_k$ (for $k > d_{\text{null}}$) correspond to eigenfunctions with greater curvature and higher oscillation frequency. The columns of $\mathbf{\Psi}_S$ define coefficient vectors for basis functions in
$\mathcal{V}_K$ that diagonalize the discrete bending-energy operator. Specifically, the $k$-th column $\boldsymbol{\psi}_{S,k} \in \mathbb{R}^K$ yields the $k$-th discrete
Ritz eigenfunction:

\begin{equation}\label{eq36}
\hat{\phi}_k(\mathbf{s})
= \sum_{j=1}^{K} \psi_{S,jk} \, \tilde{\phi}_j(\mathbf{s})
= \tilde{\boldsymbol{\phi}}(\mathbf{s})^{\top} \boldsymbol{\psi}_{S,k},
\end{equation}

where $\tilde{\boldsymbol{\phi}}(\mathbf{s})
= (\tilde{\phi}_1(\mathbf{s}), \ldots, \tilde{\phi}_K(\mathbf{s}))^{\top}$ is the vector of TPS basis functions evaluated at the spatial location $\mathbf{s}$.

\subsection{Discrete KLE with Adaptive Truncation}

Using the discrete eigenfunctions $\hat{\phi}_k$ and applying the regularization parameter $\alpha > 0$, we obtain the discrete analogue of Equation~\eqref{eq27}:

\begin{equation}\label{eq37}
K_{\alpha}(\mathbf{s}, \mathbf{s}')
\approx
\sum_{k=1}^{M}
\lambda_{k,\alpha} \,
\hat{\phi}_k(\mathbf{s}) \,
\hat{\phi}_k(\mathbf{s}'),
\end{equation}

where the discrete KLE eigenvalues (prior variances) are given by:

\begin{equation}\label{eq38}
\lambda_{k,\alpha}
= \frac{1}{1 + \alpha v_k},
\end{equation}

which approximates the continuous relationship in Equation~\eqref{eq29}, where $v_k$ are the eigenvalues of the discrete penalty matrix $\mathbf{S}$ from Equation~\eqref{eq33}. For components in the null space ($k \leq d_{\text{null}}$), we have $v_k \approx 0$, which yields $\lambda_{k,\alpha} \approx 1$. These components are therefore unpenalized and capture the low-frequency polynomial trends in the spatial field with unit prior variance. For $k > d_{\text{null}}$, we have $v_k > 0$, and $\lambda_{k,\alpha}$ decreases monotonically with increasing $k$. Larger eigenvalues $v_k$ correspond to rougher (higher-frequency) basis functions, which are increasingly shrunk by the regularization parameter $\alpha$. The decay rate is controlled by $\alpha$: larger values of $\alpha$ impose stronger regularization, causing $\lambda_{k,\alpha}$ to decay more rapidly. In practice, we truncate the expansion at $M \leq K$ components rather than using all available basis functions. We select $M$ adaptively to retain a specified proportion (typically 95-99\%) of the total variance represented by the KLE eigenvalues:

\begin{equation}\label{eq39}
M = \min\left\{m : \frac{\sum_{k=1}^{m} \lambda_{k,\alpha}}{\sum_{k=1}^{K} \lambda_{k,\alpha}} \geq \gamma \right\},
\end{equation}

where $\gamma \in (0,1)$ is the variance retention threshold (e.g., $\gamma = 0.95$). This adaptive truncation exploits the rapid decay of $\lambda_{k,\alpha}$ with increasing $k$, retaining only those components that contribute meaningfully to the covariance structure while discarding negligible high-frequency modes.

\subsection{Bayesian Interpretation}

Having established the discrete KLE representation, we now formulate the Bayesian inference framework. Consider the penalized spatial least-squares problem:

\begin{equation}\label{eq40}
\mathcal{L}(f) = \frac{1}{\sigma_\varepsilon^2} \sum_{i=1}^N (y_i - f(\mathbf{s}_i))^2 + \alpha J(f),
\end{equation}

where $y_i \overset{\mathrm{i.i.d.}}{\sim} \mathcal{N}(f(\mathbf{s}_i), \sigma_\varepsilon^2)$,
$\sigma_\varepsilon > 0$ denotes the standard deviation of the observation noise, $N$ is the number of observations at spatial locations $\mathbf{s}_i \in \mathbb{R}^d$, and $J(f) = \int_{\mathcal{D}} (\Delta^2 f)^2 \, d\mathbf{s}$ is the bending-energy functional. Let $f(\mathbf{s}) = \sum_{j=1}^K c_j \tilde{\phi}_j(\mathbf{s})$ with coefficient vector $\mathbf{c} = (c_1, \ldots, c_K)^\top$ and design matrix $\tilde{\boldsymbol{\Phi}} \in \mathbb{R}^{N \times K}$ with entries $\tilde{\Phi}_{ij} = \tilde{\phi}_j(\mathbf{s}_i)$. Then $\mathbf{y} = \tilde{\boldsymbol{\Phi}}\mathbf{c} + \boldsymbol{\varepsilon}$ with $\boldsymbol{\varepsilon} \sim \mathcal{N}(\mathbf{0}, \sigma_\varepsilon^2 \mathbf{I}_N)$, and the Gaussian likelihood is:

\begin{equation}\label{eq41}
p(\mathbf{y} \mid \mathbf{c}, \sigma_\varepsilon) = (2\pi\sigma_\varepsilon^2)^{-N/2} \exp\!\left(
-\frac{1}{2\sigma_\varepsilon^2}
\|\mathbf{y}-\tilde{\boldsymbol{\Phi}}\mathbf{c}\|^2
\right).
\end{equation}

In the Bayesian interpretation of TPS, the roughness penalty $J(f) = \mathbf{c}^\top \mathbf{S} \mathbf{c}$ corresponds to a Gaussian prior on the coefficients $\mathbf{c}$ defined through the precision matrix:

\begin{equation}\label{eq42}
\mathbf{P}_\alpha = \mathbf{I} + \alpha \mathbf{S},
\end{equation}

where $\mathbf{I}$ is the $K \times K$ identity matrix. Thus, the prior on $\mathbf{c}$ is:

\begin{equation}\label{eq43}
\pi(\mathbf{c}\mid\alpha) \propto \exp\!\left( -\frac{1}{2}\mathbf{c}^\top \mathbf{P}_{\alpha}\mathbf{c} \right),
\end{equation}

which implies $\mathbf{c} \sim \mathcal{N}(\mathbf{0}, \mathbf{K}_{\alpha})$, with covariance matrix $\mathbf{K}_{\alpha} = \mathbf{P}_{\alpha}^{-1} = (\mathbf{I} + \alpha \mathbf{S})^{-1}$. Using the eigendecomposition $\mathbf{S} = \mathbf{\Psi}_S \mathbf{V}_S \mathbf{\Psi}_S^\top$ from Equation~\eqref{eq33} and defining the transformed coefficients:

\begin{equation}\label{eq44}
\mathbf{z} = \mathbf{\Psi}_S^\top \mathbf{c},
\end{equation}

the prior transforms to:

\begin{equation}\label{eq45}
\pi(\mathbf{z}\mid\alpha)
\propto
\exp\!\left(
-\frac{1}{2}
\mathbf{z}^\top (\mathbf{I} + \alpha \mathbf{V}_S)\mathbf{z}
\right),
\end{equation}

revealing that the components are independent with diagonal covariance:

\begin{equation}\label{eq46}
z_k \sim \mathcal{N}\!\left(0,\, \lambda_{k,\alpha}\right),
\qquad
\lambda_{k,\alpha} = \frac{1}{1 + \alpha v_k},
\end{equation}

where $v_k$ are the eigenvalues of the penalty matrix $\mathbf{S}$ from Equation~\eqref{eq33}. This demonstrates that the eigendecomposition of $\mathbf{S}$ yields a diagonal prior covariance in the transformed coordinate system $\mathbf{z}$, with the KLE eigenvalues $\lambda_{k,\alpha}$ directly governing the prior variance of each component. For null space components ($k \leq d_{\text{null}}$), we have $v_k \approx 0$, yielding $\lambda_{k,\alpha} \approx 1$—these components have unit prior variance and are unpenalized. For penalized components ($k > d_{\text{null}}$), increasing $v_k$ leads to decreasing $\lambda_{k,\alpha}$, imposing stronger shrinkage on rougher basis functions. Moreover, the spatial field at any location $\mathbf{s}$ can be expressed in terms of the transformed coefficients as:

\begin{equation}\label{eq47}
f(\mathbf{s}) = \sum_{k=1}^{M} z_k \hat{\phi}_k(\mathbf{s}) = \boldsymbol{\Phi}(\mathbf{s})^\top \mathbf{z},
\end{equation}

where $\boldsymbol{\Phi}(\mathbf{s}) = (\hat{\phi}_1(\mathbf{s}), \ldots, \hat{\phi}_M(\mathbf{s}))^\top$ is the vector of discrete eigenfunctions, and we truncate at $M$ components as specified in Equation~\eqref{eq39}. At the observation locations, this gives the linear predictor $\boldsymbol{\mu} = \boldsymbol{\Phi}_{\text{obs}} \mathbf{z}$, where $\boldsymbol{\Phi}_{\text{obs}} \in \mathbb{R}^{N \times M}$ has entries $(\boldsymbol{\Phi}_{\text{obs}})_{ik} = \hat{\phi}_k(\mathbf{s}_i)$. This Bayesian formulation provides the foundation for posterior inference, where we estimate the transformed coefficients $\mathbf{z}$, the regularization parameter $\alpha$, and the noise level $\sigma_\varepsilon$ from the observed data $\mathbf{y}$. The prior structure in Equation~\eqref{eq46} allows efficient computation while maintaining the interpretability of the TPS framework, with automatic regularization through the KLE eigenvalue decay.

\subsubsection{Non-Centered Parameterization}

For computational efficiency and improved posterior geometry in MCMC sampling, we employ a non-centered parameterization \citep{papaspiliopoulos2007general,betancourt2015hamiltonian}. We introduce non-centered parameters $\tilde{z}_k \sim \mathcal{N}(0,1)$ and transform them according to:

\begin{equation}\label{eq48}
z_k = \sqrt{\lambda_{k,\alpha}}\,\tilde{z}_k = 
\begin{cases}
\tilde{z}_k, & k \leq d_{\text{null}} \\
\frac{\tilde{z}_k}{\sqrt{1 + \alpha v_k}}, & k > d_{\text{null}}
\end{cases}
\end{equation}

For null space components, the scaling factor is unity, ensuring these components remain unpenalized. For penalized components, the scaling factor $1/\sqrt{1 + \alpha v_k}$ implements the shrinkage induced by regularization. This parameterization decorrelates the latent coefficients $\{\tilde{z}_k\}$ from the hyperparameter $\alpha$, substantially improving MCMC mixing and convergence, particularly when $\alpha$ is also being estimated from the data. 

\subsubsection{Computational Implementation}

In practice, we work with truncated representations to improve computational efficiency. We truncate to $M$ components as specified in Equation~\eqref{eq39}, so that $\mathbf{z} = (z_1, \ldots, z_M)^\top$ and $\mathbf{\Psi}_S \in \mathbb{R}^{K \times M}$ contains only the first $M$ columns of the full eigenvector matrix. We pre-compute the KLE design matrix at observation locations:

\begin{equation}\label{eq49}
\boldsymbol{\Phi}_{\text{obs}} = \tilde{\boldsymbol{\Phi}} \mathbf{\Psi}_S \in \mathbb{R}^{N \times M},
\end{equation}

where $\tilde{\boldsymbol{\Phi}} \in \mathbb{R}^{N \times K}$ contains the TPS basis functions $\tilde{\phi}_j(\mathbf{s}_i)$ evaluated at observation locations, and the entries of $\boldsymbol{\Phi}_{\text{obs}}$ are $(\boldsymbol{\Phi}_{\text{obs}})_{ik} = \hat{\phi}_k(\mathbf{s}_i)$, representing the evaluation of the $k$-th discrete eigenfunction at location $\mathbf{s}_i$. The likelihood in terms of the non-centered parameters $\tilde{\mathbf{z}}$ is then:

\begin{equation}\label{eq50}
p(\mathbf{y}\mid\tilde{\mathbf{z}}, \alpha, \sigma_\varepsilon)
= (2\pi\sigma_\varepsilon^2)^{-N/2} \exp\!\left(
-\frac{1}{2\sigma_\varepsilon^2}
\|\mathbf{y}-\boldsymbol{\Phi}_{\text{obs}}\mathbf{z}(\tilde{\mathbf{z}}, \alpha)\|^2
\right),
\end{equation}

where $\mathbf{z}(\tilde{\mathbf{z}}, \alpha)$ denotes the transformation in Equation~\eqref{eq48}. 

\subsubsection{Prior Specification and Posterior Inference}

We complete the Bayesian model by assigning hyperpriors. For the observation noise standard deviation $\sigma_\varepsilon$, we use a penalized complexity (PC) prior \citep{simpson2017penalising}:

\begin{equation}\label{eq51}
\pi(\sigma_\varepsilon) = \lambda_\sigma \exp(-\lambda_\sigma \sigma_\varepsilon), \quad \lambda_\sigma = -\frac{\log(\alpha_0)}{\sigma_0},
\end{equation}

where $\sigma_0$ and $\alpha_0$ are chosen such that $P(\sigma_\varepsilon > \sigma_0) = \alpha_0$. We set $\sigma_0 = 0.5$ and $\alpha_0 = 0.05$, encoding weak prior information that the noise standard deviation is unlikely to exceed 0.5 with high probability. For the regularization parameter $\alpha$, we use a lognormal prior:

\begin{equation}\label{eq52}
\log \alpha \sim \mathcal{N}(0, \sigma_\alpha^2),
\end{equation}

with $\sigma_\alpha = 3$ to allow flexibility while maintaining weak informativeness. This prior places most mass on moderate smoothing levels while allowing the data to determine the appropriate degree of regularization. The non-centered parameters have standard normal priors: $\tilde{z}_k \sim \mathcal{N}(0,1)$ for $k=1,\ldots,M$. Thus, combining the likelihood in Equation~\eqref{eq50} with the priors, the joint posterior distribution is:

\begin{equation}\label{eq53}
p(\tilde{\mathbf{z}}, \alpha, \sigma_\varepsilon \mid \mathbf{y})
\propto
p(\mathbf{y}\mid \tilde{\mathbf{z}}, \alpha, \sigma_\varepsilon)\,
\prod_{k=1}^M p(\tilde{z}_k) \,
p(\alpha)\,
p(\sigma_\varepsilon),
\end{equation}

where the likelihood depends on $\tilde{\mathbf{z}}$ through the transformation in Equation~\eqref{eq48}.

To characterize the uncertainty in all model parameters, including $\alpha$ and $\sigma_\varepsilon$, we employ the Hamiltonian Monte Carlo (HMC) algorithm implemented in Stan \citep{carpenter2017stan, stan2018stan}, via Template Model Builder (TMB) \citep{kristensen2015tmb} and \texttt{tmbstan} \citep{monnahan2018no} packages of the statistical software R. The pre-computation of the fixed eigendecomposition in Equation~\eqref{eq33} allows for fast, linear-time calculation of the prior density and likelihood gradient in each step of the sampler, making it computationally feasible to obtain a robust representation of the full posterior distribution.

\subsubsection{The SPDE method}

For comparison purposes, we will consider the popular SPDE method \citep{lindgren2011explicit}. Briefly, instead of working directly with dense covariance matrices, the GRF is represented as solution of a stochastic partial differential equation (SPDE). By the Finite Element Method (FEM), the spatial domain is discretized, and one obtains a sparse precision matrix that approximates the covariance structure of the GRF. In simple terms, for large-scale spatial modeling purposes, this method allows for efficient computation while retaining theoretical guaranties from the Mat\'ern class of covariance families \citep{whittle1954stationary, whittle1963stochastic, lindgren2011explicit}. For a summarized explanation of this method and the prior distributions used for the hyperparameters participating in the MCMC sampling, see Appendix \ref{secA2}. 

\newpage
\section{Numerical analysis}
\label{section4}

As previously mentioned, we will refer to our proposed approach (the regularized TPS kernel) to approximate a GRF using the KLE method as regTPS-KLE. We will assess its statistical and computational performance through a series of analyzes and compare the results against the widely used SPDE method to approximate a GRF \citep{lindgren2010explicit}. For this purpose, we adopt the spatial statistical model specified in Equation \eqref{eq1} without the mean component ($\mu$); that is, no predictors were considered. The models are evaluated under four scenarios with increasing numbers of spatial locations: 50, 100, 150, and 200 points, referred to as Sce. 1, Sce. 2, Sce. 3, and Sce. 4, respectively. For each scenario, a single realization of the true GRF is simulated on the nodes of a common triangular mesh. Specifically, a covariance matrix is constructed using a Mat\'ern covariance function evaluated at the mesh nodes, and a multivariate normal sample is drawn to obtain the spatial field. The true GRF is then projected from the mesh nodes to the spatial locations and to a regular grid using projection matrices derived from the SPDE discretization. Observations are generated by adding independent Gaussian noise to the projected field at the observation locations, that is; $y_i = u(\mathbf{s}_i) + \varepsilon_i$, where $\varepsilon_i \sim \mathcal{N}(0, \sigma_\varepsilon)$. This SPDE-based framework is used consistently across all scenarios to ensure a fair comparison between the SPDE and regTPS-KLE, as both methods are fitted to the same underlying spatial field and observation process. Thus, for each scenario, we have the following (Table \ref{table:table0}):

\begin{table}[htbp]
\centering
\caption{Degrees of freedom (DoF) comparison between SPDE and regTPS-KLE approaches}
\label{table:table0}
\begin{tabular}{@{}lcccccccc@{}}
\toprule
\multirow{2}{*}{\textbf{Scenario}} & \multirow{2}{*}{Spatial Locations } & \multicolumn{3}{c}{\textbf{SPDE}} & \multicolumn{3}{c}{\textbf{regTPS-KLE}} & \multirow{2}{*}{\textbf{Ratio}} \\
\cmidrule(lr){3-5} \cmidrule(lr){6-8}
 & & $n_{\text{mesh}}$ & Hyper. & Total & $M$ & Hyper. & Total & \\
\midrule
Sce. 1 & 50  & 106 & 3 & 109 & 44  & 2 & 46  & 0.42 \\
Sce. 2 & 100 & 147 & 3 & 150 & 88  & 2 & 90  & 0.60 \\
Sce. 3 & 150 & 202 & 3 & 205 & 125 & 2 & 127 & 0.62 \\
Sce. 4 & 200 & 247 & 3 & 250 & 170 & 2 & 172 & 0.69 \\
\bottomrule
\multicolumn{9}{@{}l@{}}{\footnotesize Hyper. = Hyperparameters; Ratio = $M/n_{\text{mesh}}$ (spatial DoF ratio)} \\
\multicolumn{9}{@{}l@{}}{\footnotesize SPDE hyperparameters: $\sigma_\varepsilon, \sigma_u, \rho$; regTPS-KLE: $\sigma_\varepsilon, \alpha$} \\
\end{tabular}
\end{table}

Finally, the mathematical formulation of the models can be found in Appendix \ref{secA3} and the R code to simulate the data in Appendix \ref{secA4}. 

\subsection{Evaluations of the regTPS-KLE}

\begin{figure}[!ht]
\centering
\includegraphics[width=13cm, height=10cm]{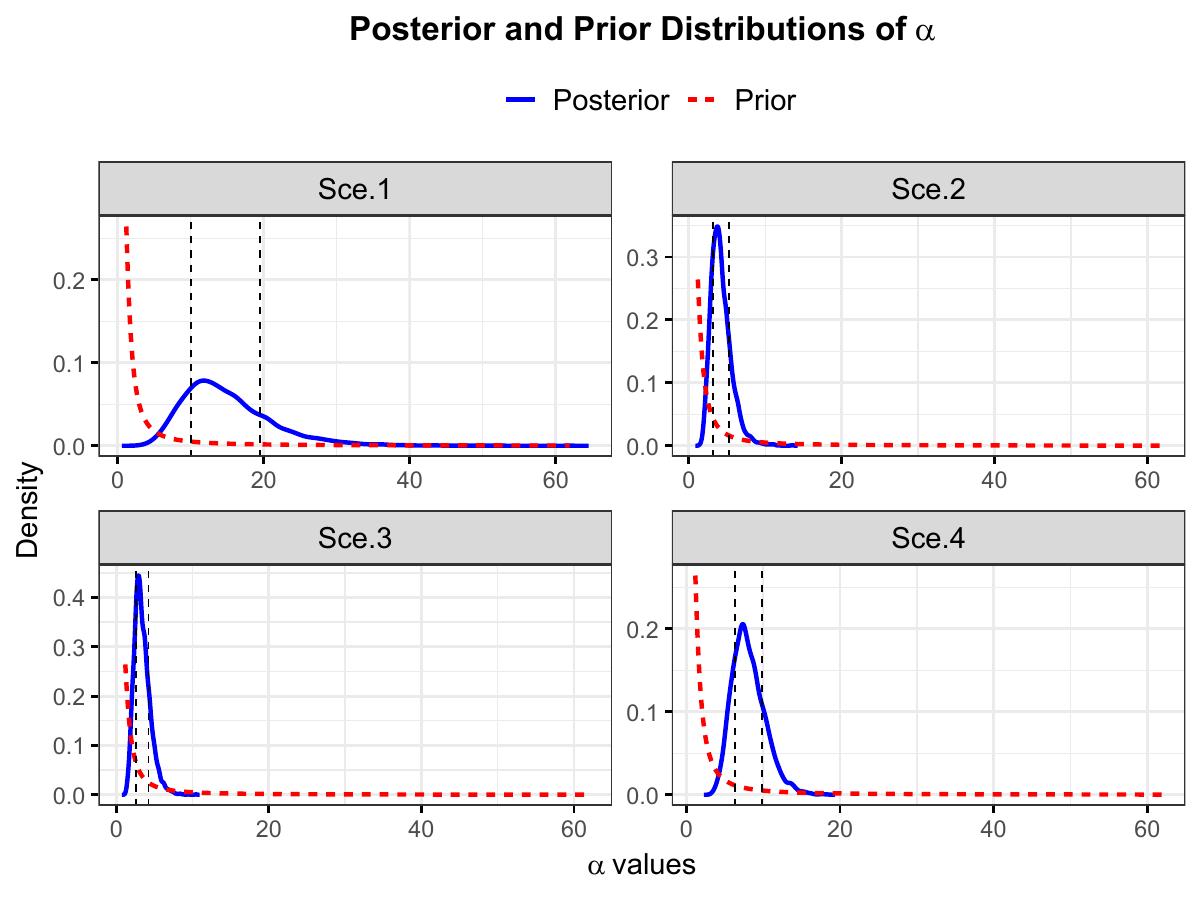}
\caption{Posterior (blue color) and prior distributions (red color) for the penalty parameter $\alpha$. The vertical dashed lines indicates the quantiles (0.2 and 0.8) associated with the posterior distributions.}\label{fig:fig1}
\end{figure}

Figure  \ref{fig:fig1} shows the prior and posterior distribution of $\alpha$ (penalty parameter). In this case, for the simulated data, the posterior distribution in all scenarios (Sce. 1 to Sce. 4) are narrower than before, meaning that the data provided much more information than previously, allowing us to compute the value of $\alpha$ accordingly, since it could normally be a problematic parameter in the sampling estimation process. 

\newpage
To assess whether the posterior distribution is coherent with the prior structure and to diagnose potential identifiability issues, we examine the relationship between prior standard deviations and posterior means of the basis coefficients. For each retained component $k = 1, \ldots, M$, we compute :

\begin{itemize}
\item \textbf{Prior standard deviation}: $\text{SD}_{\text{prior}}(z_k) = \sqrt{\lambda_{k,\alpha}}$, where $\lambda_{k,\alpha} = (1 + \alpha v_k)^{-1}$.
\item \textbf{Posterior mean (absolute value)}: $|\bar{z}_k| = |\mathbb{E}[z_k \mid \mathbf{y}]|$, estimated from MCMC samples.
\end{itemize}

\begin{figure}[!ht]
\centering
\includegraphics[width=13cm, height=13cm]{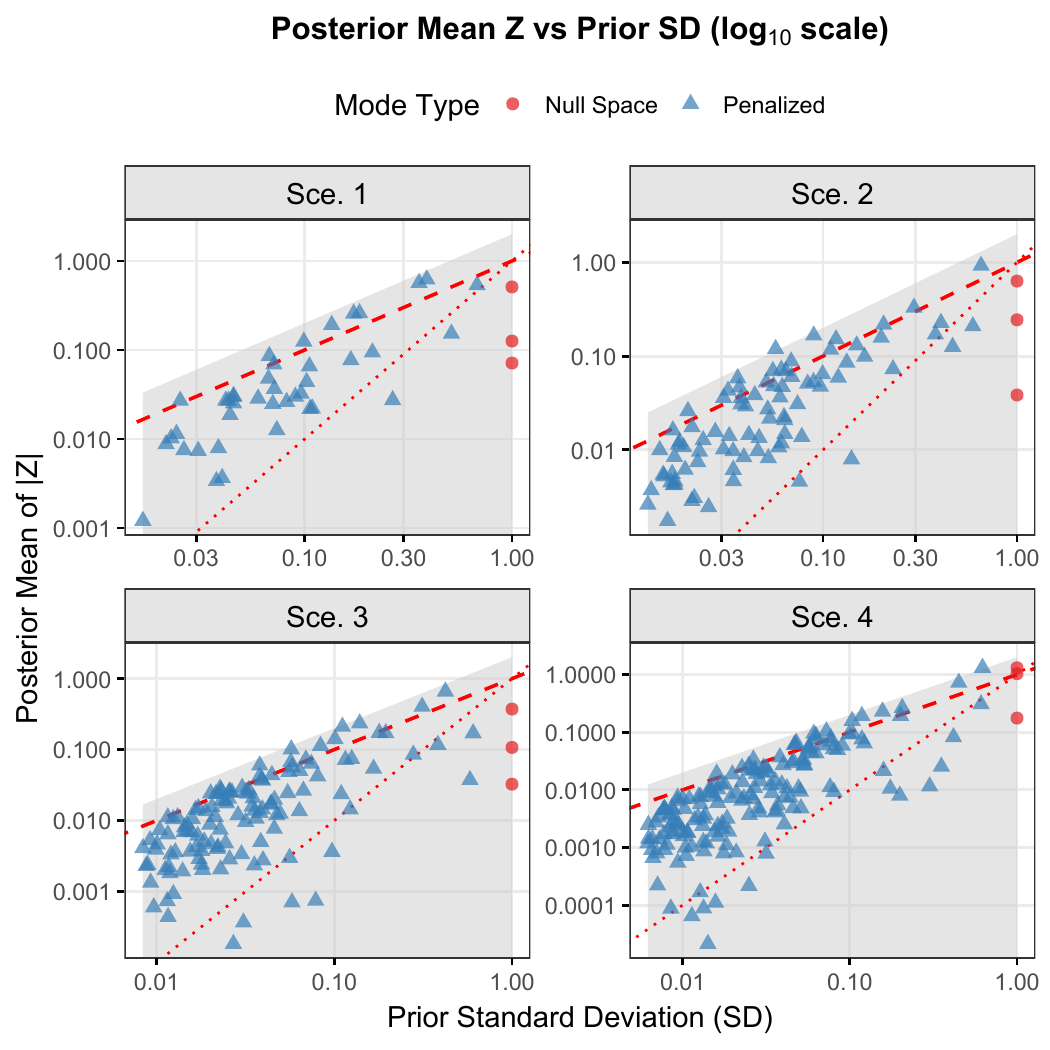}
\caption{Posterior means $|\mathbb{E}[z_k \mid \mathbf{y}]|$ versus prior standard deviations $\sqrt{\lambda_{k,\alpha}}$ in a $\log_{10}$ scale for all scenarios. The red circles indicates the null space modes ($k \leq 3$) with unit prior variance and blue triangles indicates the penalized modes with $\lambda_{k,\alpha} = (1 + \alpha v_k)^{-1}$. The dashed line is the theoretical prior expectation $\mathbb{E}[|z_k|] = \sqrt{2\lambda_{k,\alpha}/\pi}$ while the gray region is $\pm 2\text{SD}$ band.}\label{fig:fig2}
\end{figure}

Figure \ref{fig:fig2} plots $|\bar{z}_k|$ against $\text{SD}_{\text{prior}}(z_k)$ for all scenarios in a $\log_{10}$ scale, where the dashed line represents the theoretical prior relationship $\mathbb{E}[|z_k|] = \text{SD}_{\text{prior}}(z_k) / \sqrt{\pi/2}$, and the shaded region indicates a $\pm 2\text{SD}$ credible band. Points falling along the theoretical line suggest that the posterior is dominated by the prior (weak data information), while deviations indicate data-driven learning. The clear separation between null space modes (unit prior SD) and penalized modes (decaying prior SD) confirms that the eigenvalue spectrum correctly encodes the smoothness hierarchy of the spatial field. It verifies that the prior standard deviations computed from the discrete eigenvalue problem correctly approximate the continuous theory and that the MCMC sampler successfully explores the posterior without numerical artifacts.

\begin{figure}[!ht]
\centering
\includegraphics[width=13cm, height=10cm]{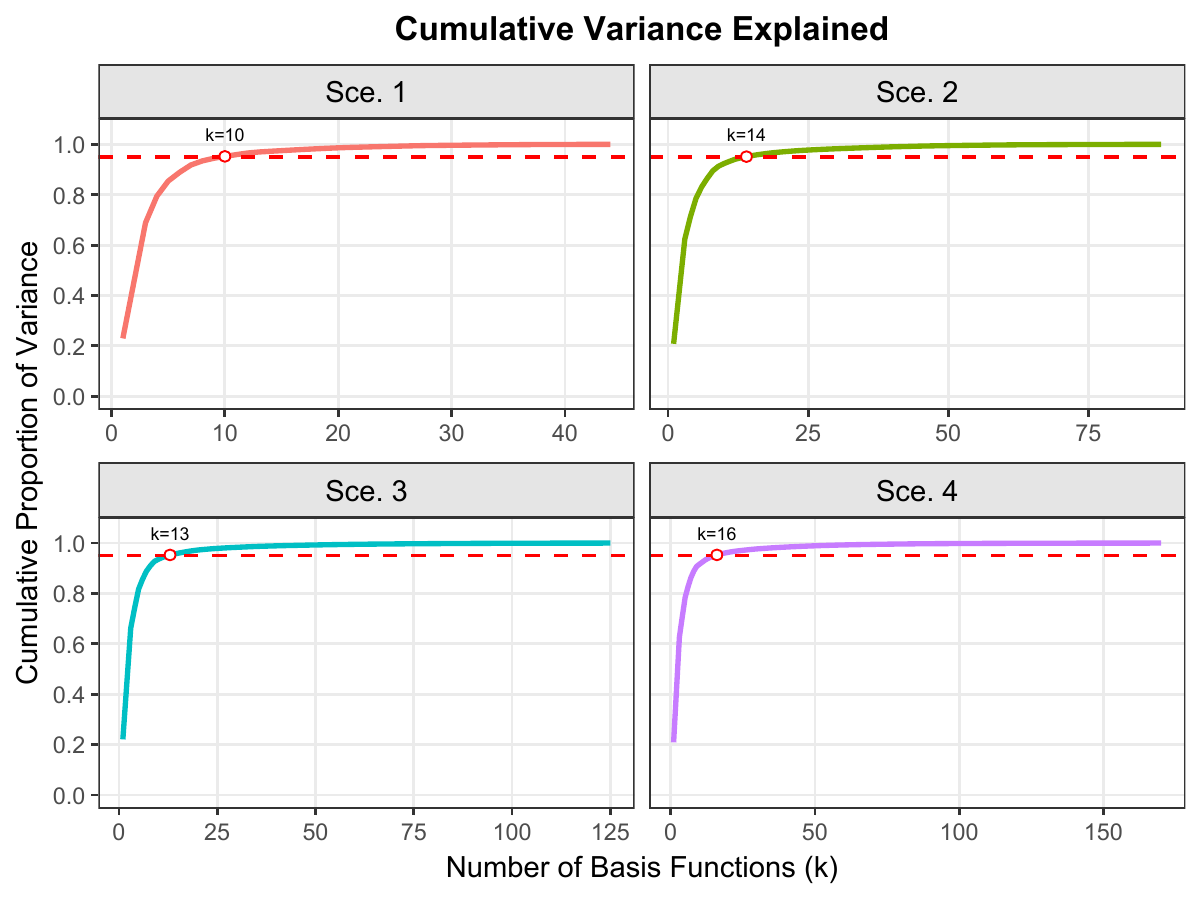}
\caption{Cumulative proportion of variance explained by the truncated KLE. The dashed horizontal line indicates the 99\% threshold used to select the truncation level.}\label{fig:fig3_b}
\end{figure}

\begin{figure}[!ht]
\centering
\includegraphics[width=13cm, height=10cm]{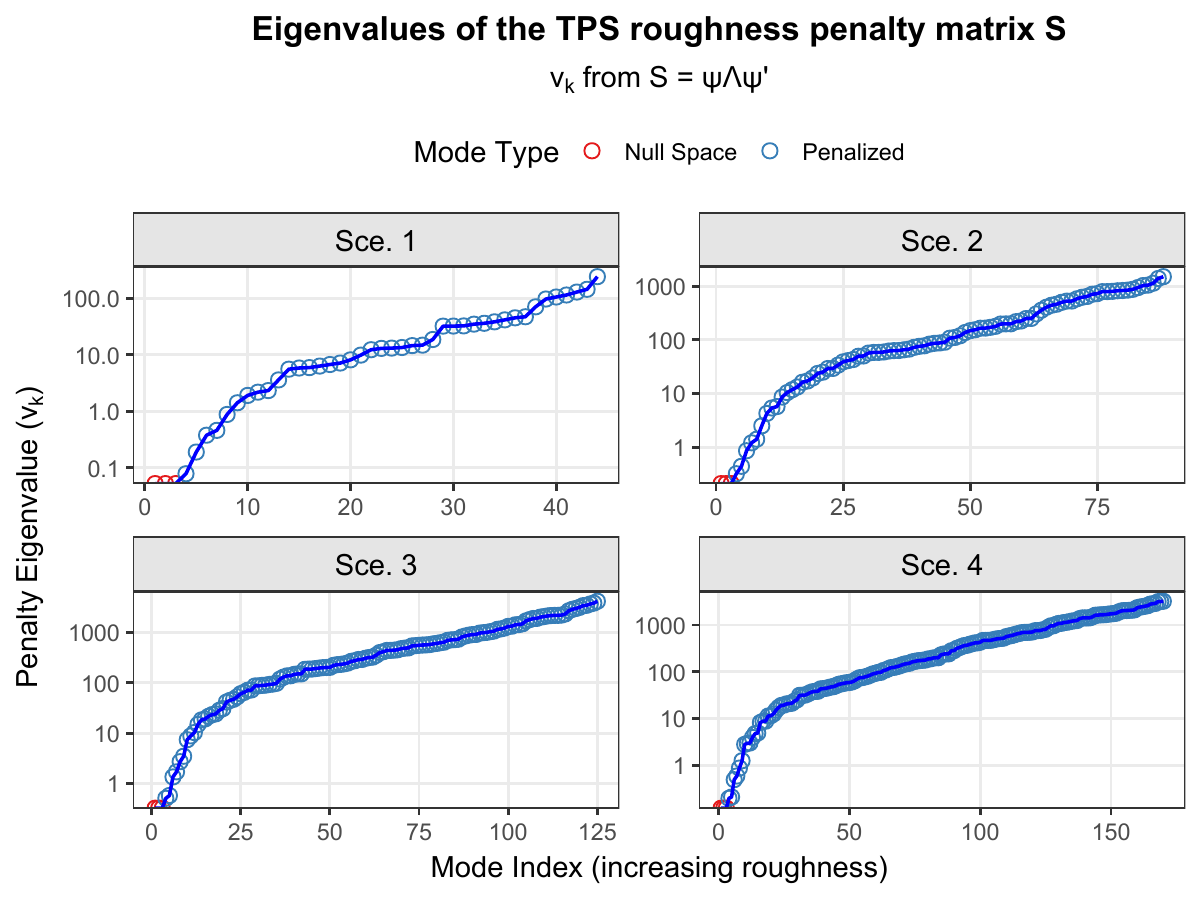}
\caption{The penalty eigenvalues $v_k$ increase rapidly after the null-space components, reflecting increasing roughness of higher-order TPS basis functions. Small values of $v_k$ correspond to smooth, low-frequency modes, whereas large values penalize highly oscillatory components}\label{fig:fig3_c}
\end{figure}

\newpage
Since the truncation point should depend on the required level of accuracy in reconstructing the covariance function, Figure  $\ref{fig:fig3_b}$ shows the cumulative variance explained by the components of the regTPS-KLE model. As shown, 99\% of the total variance is explained using fewer components than the number of basis functions, confirming that the model effectively reduces the dimensionality of the spatial data (with the curves rapidly approaching 99\% of explained variance). Selecting a lower threshold of explained variance would further improve computational efficiency; however, in this case, the number of estimated TPS basis functions does not approximate to the number of mesh nodes derived from the discretized spatial domain in the SPDE approach for a fair comparison. Figure \ref{fig:fig3_c} shows the eigenvalue spectrum of the $\mathbf{S}$ penalty matrix. The eigenvalues increase rapidly after the null-space components, reflecting the growing roughness associated with higher-order basis functions. This behavior is expected, as the penalty operator assigns small eigenvalues to smooth, low-frequency modes and increasingly large eigenvalues to highly oscillatory components. While the penalty eigenvalues grow with increasing roughness, the corresponding KLE eigenvalues decrease monotonically, leading to fast saturation of the cumulative variance explained (Figure  \ref{fig:fig3_b}). These results collectively demonstrate that the regTPS-KLE model concentrates most of the spatial variability in a small number of smooth modes.

\clearpage
\subsection{SPDE and regTPS-KLE comparisons}

\begin{figure}[!ht]
\centering
\includegraphics[width=16cm, height=16cm]{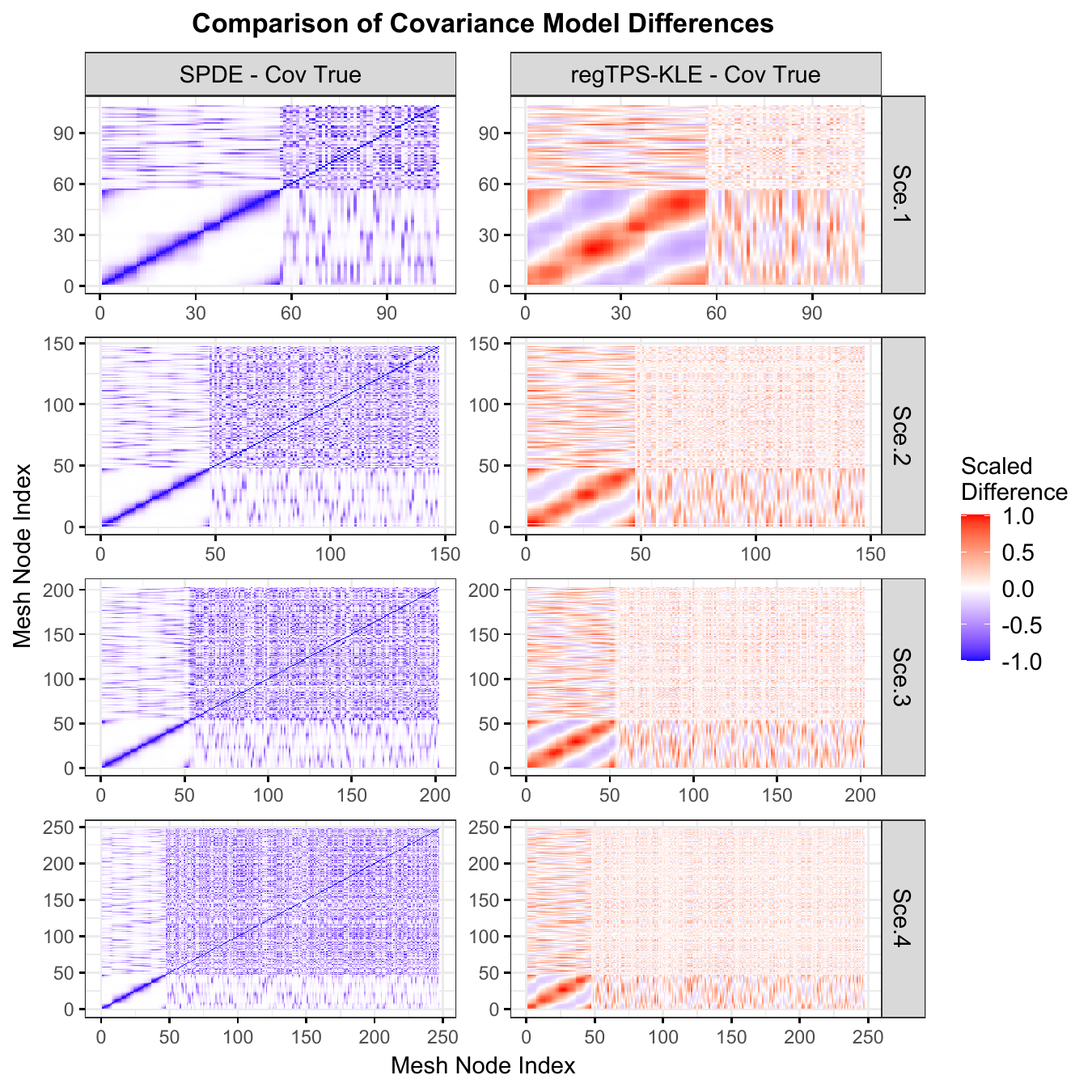}
\caption{Normalized estimation errors of the SPDE (left side) and regTPS-KLE model (right side) relative to the true Mat\'ern covariance. Spatial patterns of the heatmap reveal underestimation (blue) and overestimation (red) for each model.}\label{fig:fig5}
\end{figure}

Since covariance matrices obtained from the SPDE and regTPS-KLE methods are defined in different spaces, they cannot be compared directly. Thus, to enable a fair comparison, we project them into a common space. Specifically, the covariance matrix $\mathbf{K}_\alpha$ from regTPS-KLE is mapped from its low-dimensional basis function space to the high-dimensional mesh node space (the mesh built for the SPDE method). For example, according to the variance explained for \textbf{Sce. 1}, only 44 basis functions are needs to build the approximated GRF; hence we evaluate these basis functions at the number of mesh node locations coming from the discretized domain using the SPDE method (106 nodes). Essentially, we are building the spatial field at the mesh nodes from the basis functions coefficients. After this, the difference between the true covariance matrix and the obtained from the SPDE and regTPS-KLE methods were plotted (Figure  \ref{fig:fig5}). Although there is a small positive difference between the values of the true and regTPS-KLE covariance matrices, as we expect, the patterns of the points are very similar for all the scenarios. On the contrary, the covariance matrix estimated for the SPDE method gives us negative difference values compared with the true covariance matrix. 

\begin{figure}[!ht]
\centering
\includegraphics[width=19cm, height=16cm]{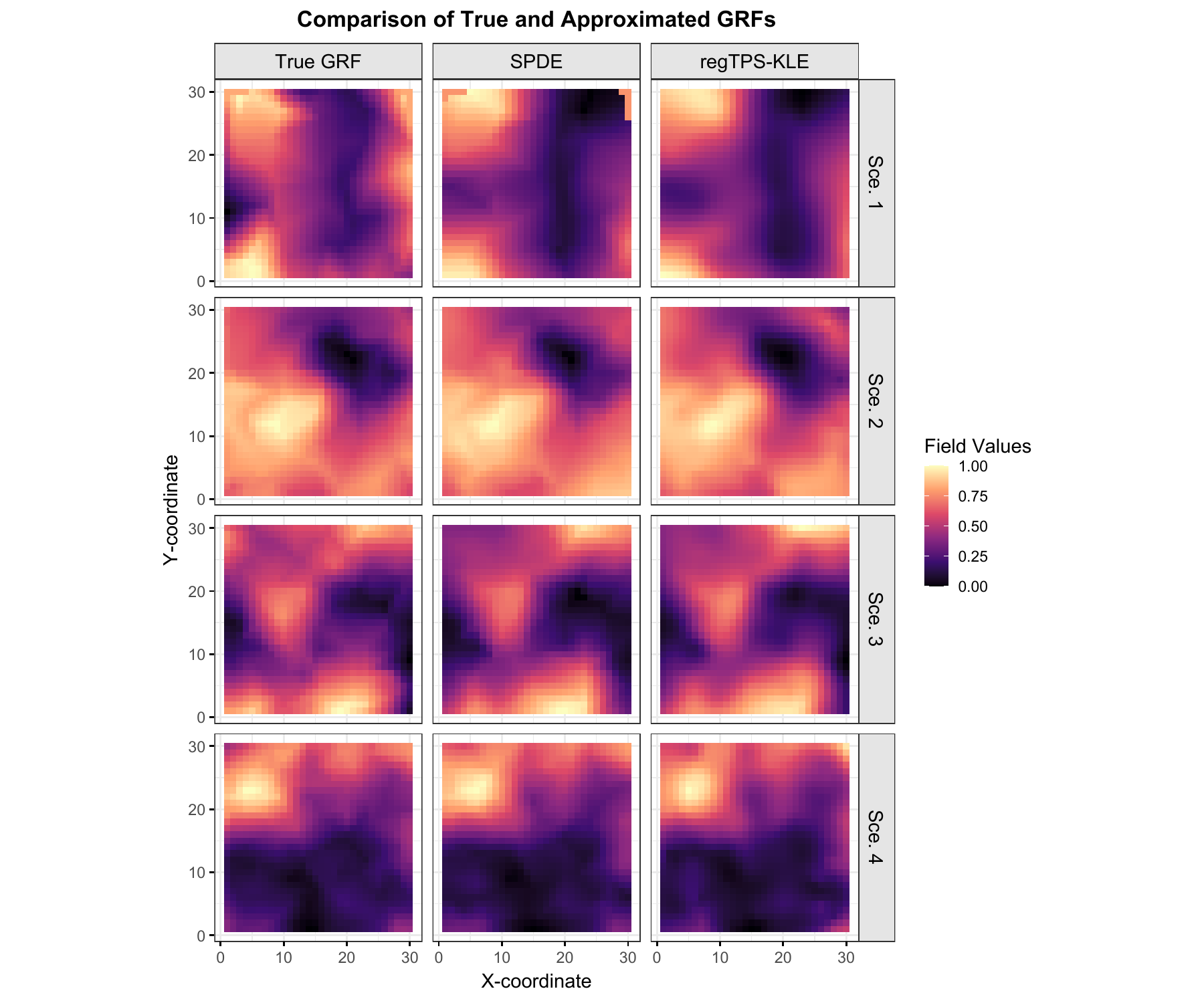}
\caption{Posterior medians of the true GRF and the approximated GRF by the SPDE and regTPS-KLE methods across all scenarios.}\label{fig:fig6}
\end{figure}

Table \ref{table:table1} summarizes the common metrics used to evaluate the predictive performance of a spatial statistical model. As expected, the SPDE approach outperforms the regTPS-KLE models. This is because the data were generated under a Mat\'ern covariance structure, for which the SPDE method has a direct representation. Nevertheless, the statistical performance difference between the SPDE and regTPS-KLE models is relatively small. In fact, for the root mean square error (RMSE) and mean absolute error (MAE) metrics, the discrepancy between the true GRF and the predictions from regTPS-KLE is close to zero. Figure  \ref{fig:fig6} presents the interpolation of the approximated GRF obtained with the SPDE and regTPS-KLE models and their comparison with the true GRF. Across all scenarios, both approaches are able to correctly reproduce the underlying spatial structure of the true spatial field. This holds even when the number of spatial observations is relatively small (for example, Sce. 1 only considers 50 spatial observations associated with their respective 50 spatial locations).

\begin{table}[!h]
\centering
\caption{Comparison of SPDE and regTPS-KLE models with the true GRF using a Mat\'ern covariance function across scenarios.}
\label{table:table1}
\centering
\begin{tabular}[t]{llrrr}
\toprule
\textbf{Scenario} & \textbf{Method}  & \textbf{RMSE} & R$^2$ & \textbf{MAE}\\
\midrule
 & SPDE & 0.343 & 0.650 & 0.219\\
\multirow[t]{-2}{*}{\raggedright\arraybackslash Sce. 1} & regTPS-KLE & 0.391 & 0.668 & 0.264\\
\addlinespace
\midrule
 & SPDE & 0.165 & 0.972 & 0.117\\
\multirow[t]{-2}{*}{\raggedright\arraybackslash Sce. 2} & regTPS-KLE & 0.210 & 0.958 & 0.167\\
\addlinespace
\midrule
 & SPDE & 0.187 & 0.916 & 0.137\\
\multirow[t]{-2}{*}{\raggedright\arraybackslash Sce. 3} & regTPS-KLE & 0.209 & 0.914 & 0.149\\
\addlinespace
\midrule
 & SPDE & 0.145 & 0.981 & 0.114\\
\multirow[t]{-2}{*}{\raggedright\arraybackslash Sce. 4} & regTPS-KLE & 0.181 & 0.970 & 0.139\\
\bottomrule
\end{tabular}
\end{table}

\begin{figure}[!ht]
\centering
\includegraphics[width=15cm, height=15cm]{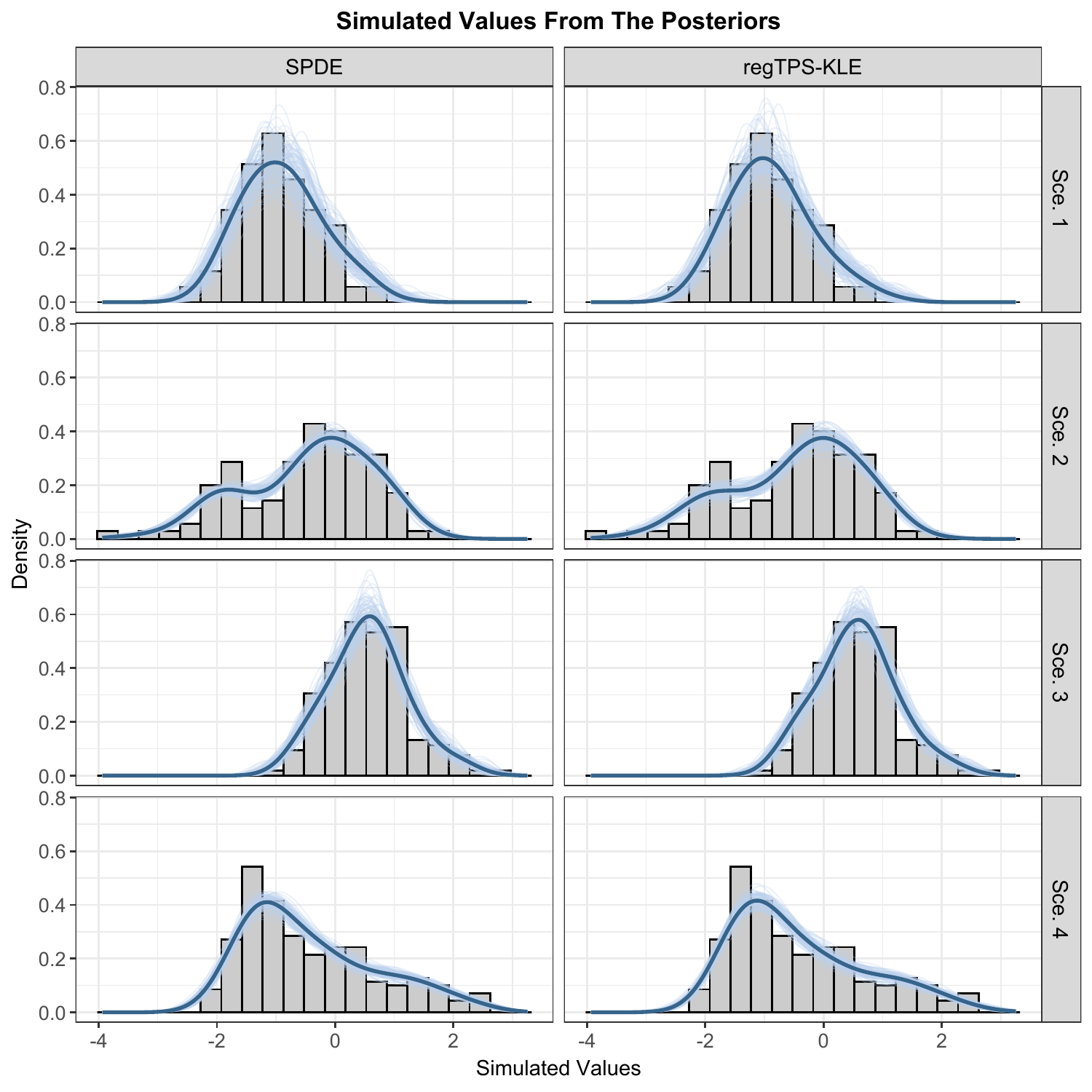}
\caption{Observed values (histogram) and 100 posterior predictive distributions (light blue lines) from the SPDE and regTPS-KLE methods for Sce.1 to Sce. 4. The ``dark" blue line is the posterior mean of all the posterior predictive distributions.}\label{fig:fig7}
\end{figure}

\newpage
Figure  \ref{fig:fig7} shows the posterior predictive distributions for both approaches. This allows us to evaluate the complete probabilistic description of model predictions, rather than limiting inference to point estimates. By incorporating both process and measurement uncertainty, the predictive posterior allows for rigorous quantification of uncertainty around predictions. From the figure, we can see that both methods can correctly reproduce new simulated data from the posterior distribution obtained for the parameters in the models for all scenarios.

\newpage
As an additional comparison based on the RMSE, R$^{2}$ and MAE metrics, we simulated a GRF using the exponential covariance function (Section \ref{subsection2.2}) and applied the same procedure described in Section \ref{section4} to generate the observed data. The results, summarized in Table \ref{table:table2}, show that in one out of four scenarios, the regTPS-KLE outperforms the SPDE models (Sce. 2). This suggests that regTPS-KLE could be effectively extended beyond the Mat\'ern correlation to other covariance functions while still maintaining a well-behaved  predictive performance.

\begin{table}[!ht]
\centering
\caption{Comparison of SPDE and regTPS-KLE models with the true GRF using a Exponential covariance function across scenarios.}
\label{table:table2}
\begin{tabular}[t]{llrrr}
\toprule
\textbf{Scenario} & \textbf{Method}  & \textbf{RMSE} & R$^2$ & \textbf{MAE}\\
\midrule
 & SPDE & 0.298 & 0.868 & 0.201\\
\multirow[t]{-2}{*}{\raggedright\arraybackslash Sce. 1} & regTPS-KLE & 0.357 & 0.819 & 0.274\\
\addlinespace
\midrule
 & SPDE & 0.261 & 0.933 & 0.210\\
\multirow[t]{-2}{*}{\raggedright\arraybackslash Sce. 2} & regTPS-KLE & 0.257 & 0.930 & 0.211\\
\addlinespace
\midrule
 & SPDE & 0.213 & 0.925 & 0.164\\
\multirow[t]{-2}{*}{\raggedright\arraybackslash Sce. 3} & regTPS-KLE & 0.270 & 0.906 & 0.205\\
\addlinespace
\midrule
 & SPDE & 0.263 & 0.927 & 0.199\\
\multirow[t]{-2}{*}{\raggedright\arraybackslash Sce. 4} & regTPS-KLE & 0.370 & 0.860 & 0.270\\
\bottomrule
\end{tabular}
\end{table}

\begin{figure}[!ht]
\centering
\includegraphics[width=14cm, height=10.5cm]{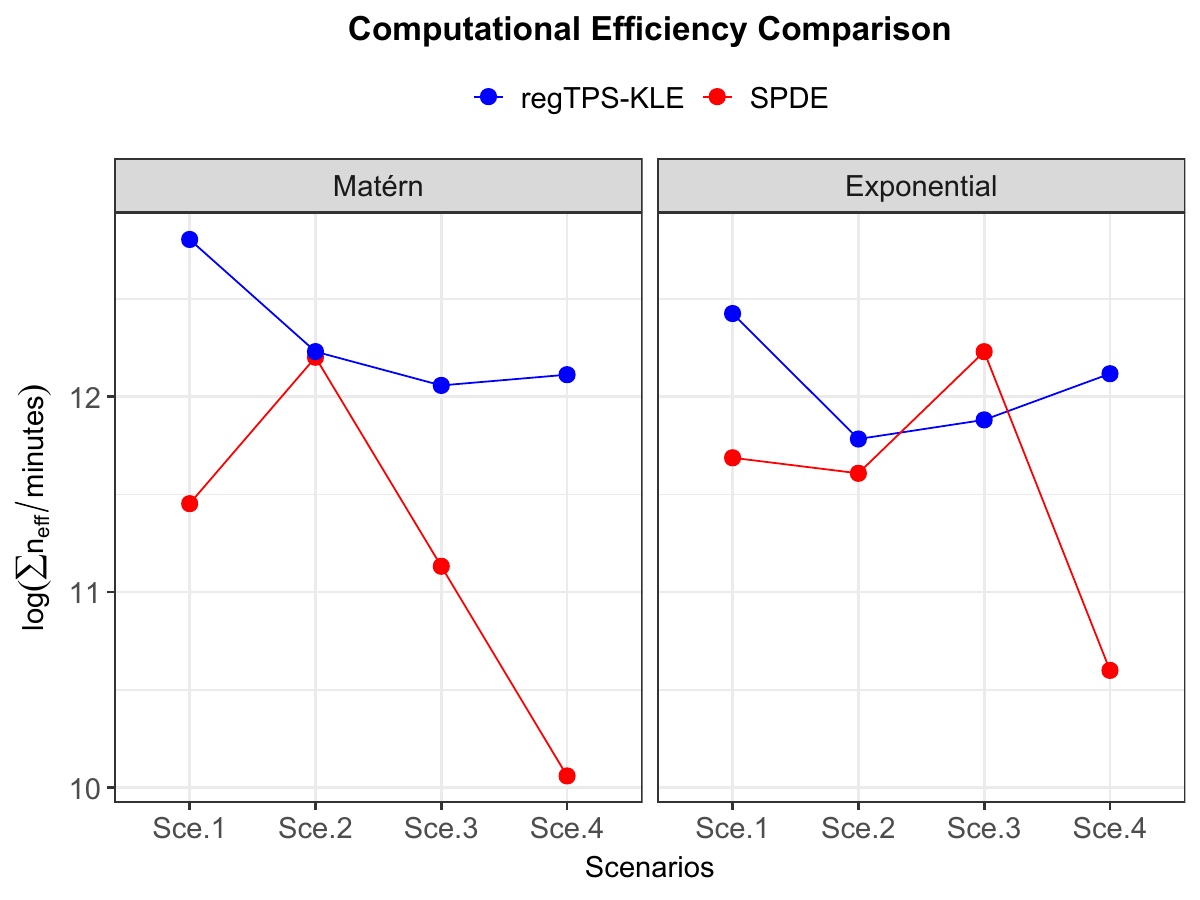}
\caption{Computational efficiency $\log\left(\frac{\sum \text{n}_\text{eff}}{\text{minutes}}\right)$ calculated for the different scenarios in the SPDE and regTPS-KLE models.}\label{fig:fig8}
\end{figure}

\newpage
Lastly, Figure  \ref{fig:fig8} presents the computational efficiency of the SPDE and regTPS-KLE models for all the scenarios. Computational efficiency is measured as the number of effective samples generated per unit time ($\log\left(\frac{\text{ESS}}{\text{time}}\right) = \log\left(\frac{\sum \text{n}_\text{eff}}{\text{time}}\right)$), where ``time'' denotes the number of minutes required for the MCMC chains to reach convergence. As shown, regTPS-KLE is more computationally efficient than the SPDE models in most scenarios. Only in scenario 2 (Mat\'ern covariance function) and scenario 3 (exponential covariance function) do the SPDE models exhibit comparable or superior efficiency. In the remaining scenarios, regTPS-KLE consistently demonstrates superior computational performance.

\clearpage
\section{Real Application: NO\texorpdfstring{$_{2}$}{} data}
\label{section5}

\begin{figure}[!ht]
\centering
\includegraphics[width=15cm, height=10cm]{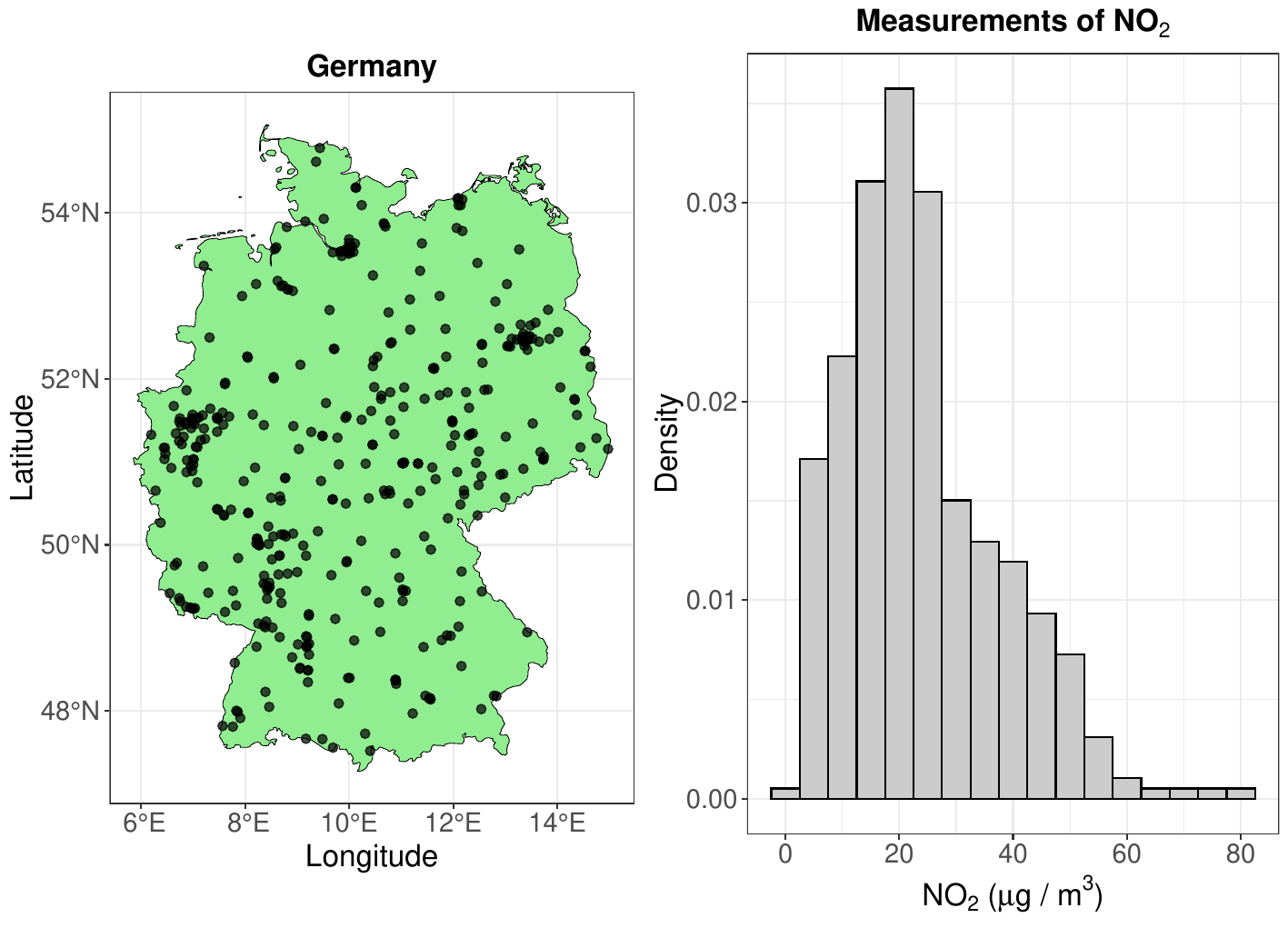}
\caption{Geographical distribution of ground stations in Germany and histogram for NO$_{2}$ measurements used in this study.}\label{fig:fig9}
\end{figure}

For our analysis, we use the concentration measurements from 2017 collected national ground monitoring stations in Germany and Netherlands (416 and 66 stations respectively). The original hourly data were obtained from the European Environment Agency (EEA). Negative values were treated as missing; an annual average concentrations were calculated by taking the mean of available hourly measurements while omitting the missing values. As shown in Figure  \ref{fig:fig9}, the observed data are not Gaussian distributed; therefore, a square root transformation was applied to have an approximate Gaussian distribution. Furthermore, \citet{lu2023comparison} showed that a Gaussian likelihood provides the best fit to the observed data. For our analysis, we focus only on the German dataset, given the higher number of monitoring stations compared with the Netherlands. A detailed description of the data can be found in \citep{lu2020evaluation}. 

\subsection{Results for the SPDE and regTPS-KLE approaches applied to NO\texorpdfstring{$_{2}$}{} data}

\begin{figure}[!ht]
\centering
\includegraphics[width=14cm, height=8.5cm]{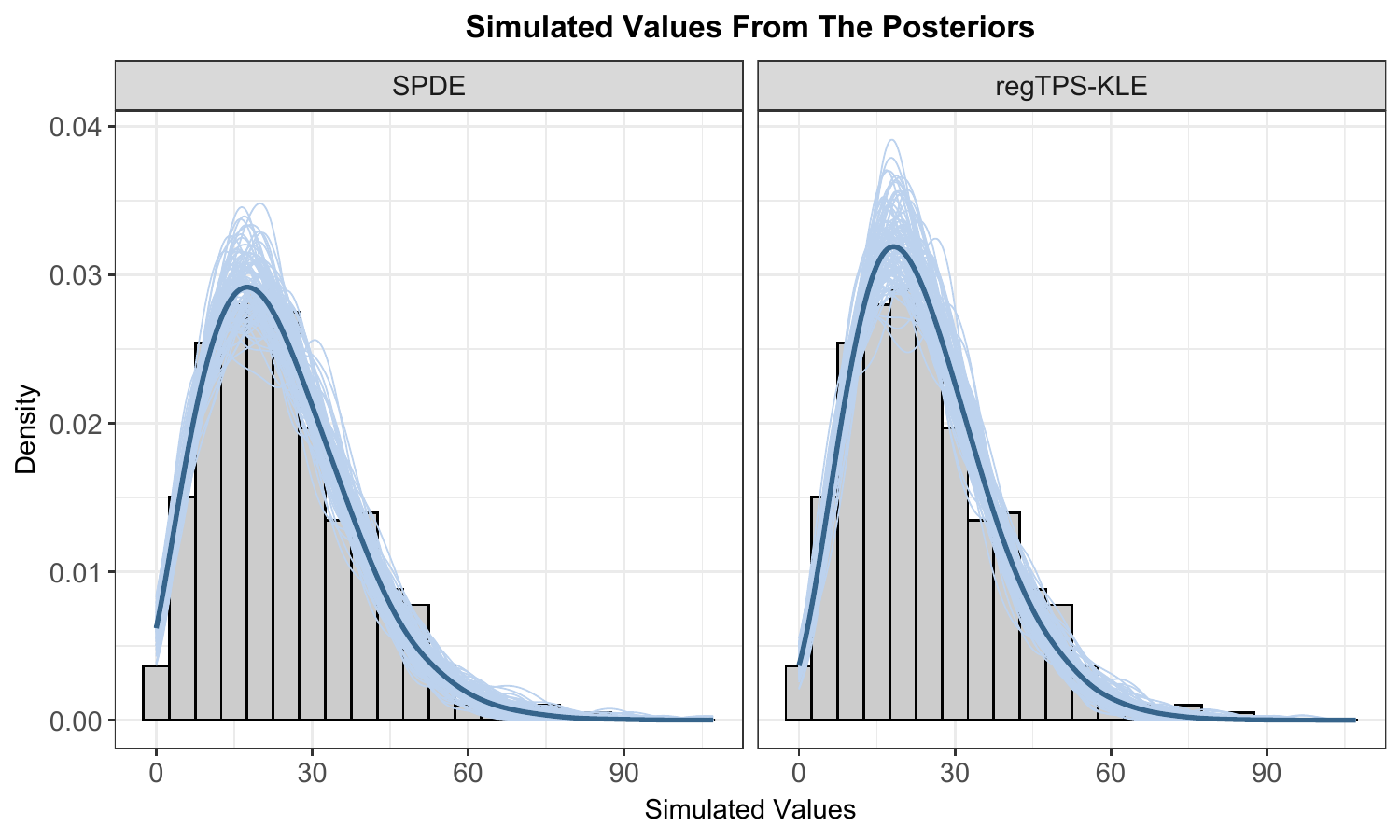}
\caption{Observed data (histogram) and 100 posterior predictive distributions (light blue lines) from the SPDE and regTPS-KLE methods applied to the NO$_{2}$ pollution in Germany. The ``dark" blue line is the posterior mean of all the posterior predictive distributions.}\label{fig:fig10}
\end{figure}

Both models quickly converged across all MCMC chains. While PC priors for SPDE models enable a simple hyperprior specification, MCMC sampling remains challenging for the SPDE method. In this case, the parameters $\rho$ and $\sigma_u$ used a Fréchet and exponential prior distributions, respectively, following the suggestion of \citep{cavieres2024approximated}. However, the chains convergence also is influenced by the mesh configuration used to discretize the spatial domain. In contrast, the regTPS-KLE approach is simpler to implement. The only parameter that could cause convergence problems is $\alpha$. For this analysis, assigning a lognormal prior distribution to $\alpha$ is sufficient to ensure reliable convergence. Regarding predictive performance, the SPDE model successfully predicted new observations drawn from the posterior, confirming its suitability for modeling the observed NO$_{2}$ data. The regTPS-KLE model can also correctly represent the NO$_{2}$ data, but with consistently higher posterior predictive values compared to the SPDE model (Figure  \ref{fig:fig10}). 

\begin{figure}[!ht]
\centering
\includegraphics[width=16cm, height=17cm]{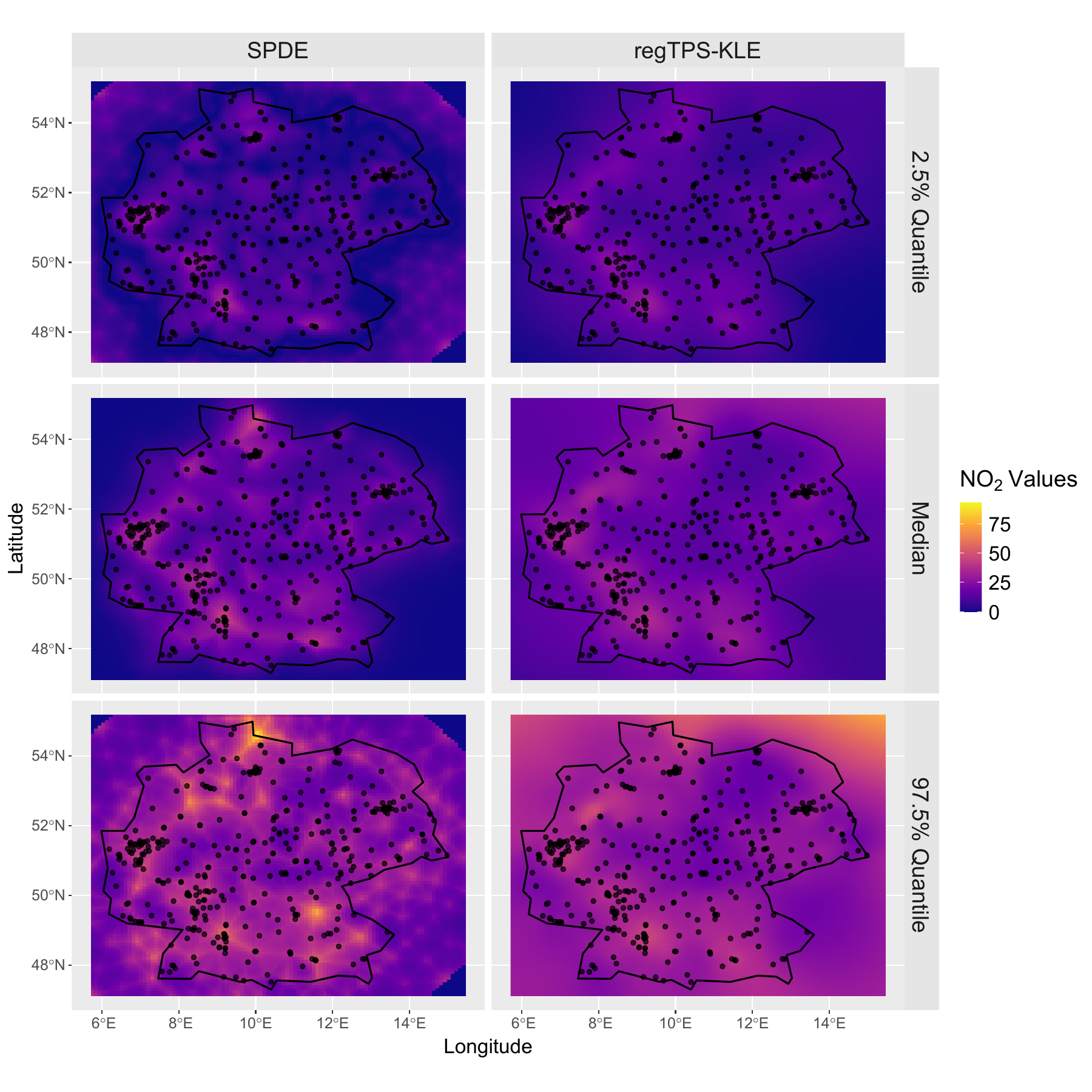}
\caption{Posterior median of NO$_{2}$ predictions/interpolations, and lower and upper quantiles (2.5\% and 97.5\% respectively) in Germany for the SPDE and regTPS-KLE models.}\label{fig:fig11}
\end{figure}

\newpage
The SPDE and regTPS-KLE models were applied to predict/interpolate the NO$_{2}$ concentrations across Germany (Figure  \ref{fig:fig11}). Both models exhibit comparable large-scale spatial patterns, particularly in the lower quantile and posterior median, where most of the average pollution in the regions is consistently identified. However, the SPDE model shows a tendency to overestimate the NO$_{2}$ concentrations in the posterior median, which may reflect its spatial smoothing properties. However, there is a high discrepancy in the upper quantiles since the SPDE model predicts elevated concentrations in the north and southeast of Germany, whereas the regTPS-KLE model attributes the highest concentrations values to the western regions. For the above, to evaluate the predictive performance of the SPDE and regTPS-KLE models, under MCMC sampling, we used the leave-one-out cross validation criterion (loo-cv, \cite{vehtari2017practical}). It is a technique that evaluates predictive performance by systematically omitting each observation from the data (NO$_{2}$) and assessing how well the model predicts the held-out value. Essentially, its use Pareto-smoothed importance sampling (PSIS-LOO), which re-weights posterior draws to approximate the leave-one-out distribution providing an efficient estimate of out-of-sample predictive accuracy. As we can see, regTPS-KLE outperforms the SPDE method for NO$_{2}$ models in predictive terms. It achieves the highest expected log predictive density (elpd), serving as the reference for comparison. Therefore, relative to this reference point (the regTPS-KLE model), the SPDE model shows a $\Delta$ELPD of -26.4 with a standard error of 7.3, indicating an important and statistically significant reduction in predictive accuracy (Table \ref{table:table3}). Our analysis also revealed that both approaches performed with high computational efficiency; however, the regTPS-KLE model demonstrated a marginal advantage over the SPDE framework. Specifically, the regTPS-KLE model achieved a computational efficiency of 12.98, compared to 12.53 for the SPDE model. In simple words, regTPS-KLE produces effective samples at a faster rate than the SPDE model with less number of basis functions (Table \ref{table:table3}).

\newpage

\begin{table}[ht]
\centering
\caption{LOO-CV model comparison between the SPDE and regTPS-KLE approaches. $\Delta$ELPD = difference in expected log predictive density relative to best model; SE = standard error of the difference; Computational Efficiency $\log\left(\frac{\sum \text{n}_\text{eff}}{\text{minutes}}\right)$; Basis indicates number of basis functions ($M$) for regTPS-KLE and mesh nodes ($n_\text{mesh}$) for SPDE.}
\label{table:table3}
\begin{tabular}{lccccr}
\hline
Model & $\Delta$ELPD & SE & Time (min) & Computational Efficiency & Basis\\
\hline
regTPS-KLE & \textbf{0.0} & --- & 6.11 & 12.94 & $M = 207$ \\
SPDE & $-26.4$ & 7.3 & 6.58 & 12.53 & $n = 300$\\
\hline
\end{tabular}
\end{table}

\clearpage
\section{Discussion}
\label{section6}

Considering all the previous analyses and comparisons between the SPDE and regTPS-KLE approaches, it is important to recognize their fundamental differences in modeling assumptions. The regTPS-KLE approach, while providing strong theoretical development and offering computational advantages through eigendecomposition, represents a restrictive subclass of GRFs compared to the SPDE framework. A key distinction lies in the treatment of spatial smoothness. The SPDE method provides flexibility in controlling the smoothness parameter $\nu$ of the Mat\'ern covariance family \citep{whittle1963stochastic, stein1999interpolation}, which directly governs the mean-square differentiability of the spatial random field. Specifically, a GRF with Mat\'ern covariance is $\lfloor \nu \rfloor$ times mean-square differentiable \citep{williams2006gaussian, gneiting2010matern}. By varying $\nu$, practitioners can accommodate spatial processes ranging from rough, non-differentiable fields ($\nu = 0.5$, corresponding to the exponential covariance) to infinitely smooth fields ($\nu \to \infty$, approaching the squared exponential covariance). More generally, the connection between SPDE operators and Mat\'ern smoothness is given by the relationship $\nu = \ell - d/2$, where $\ell$ denotes the order of the differential operator and $d$ is the spatial dimension \citep{lindgren2011explicit}. 
For the biharmonic operator in two dimensions, we have $\ell = 2$ (since $\Delta^2$ is a fourth-order operator involving second derivatives), yielding $\nu = 2 - 2/2 = 1$ for the standard TPS formulation \citep{wahba1990spline}. 
While the regularization parameter $\alpha$ controls the effective spatial range (correlation length) of the field, with larger $\alpha$ inducing stronger smoothing and longer-range spatial dependence, it cannot alter the fundamental smoothness class determined by the operator order. To achieve different levels of differentiability within the TPS framework, one would need to change the order of the differential operator itself (e.g., using a higher-order operator such as $\Delta^3$ in two dimensions to obtain $\nu \approx 2$), though this would require reconfiguring the entire basis and penalty structure \citep{duchon1977splines, meinguet1979multivariate}. In this way, the smoothness constraint has both advantages and limitations. On the positive side, fixing $\nu$ reduces the number of hyperparameters to estimate and eliminates potential identifiability issues that can arise when jointly estimating smoothness and range parameters in the Mat\'ern model \citep{zhang2004inconsistent, stein2004approximating}. The once mean-square differentiable smoothness class $\nu = 1$ is often appropriate for many environmental and physical processes, providing a reasonable default that balances flexibility with computational tractability \citep{cressie1989geostatistics, stein1999interpolation}. However, the fixed smoothness assumption may be restrictive for applications where the true spatial process exhibits markedly different regularity properties, such as processes with sharp discontinuities or, conversely, phenomena requiring substantially smoother representations.
Another important consideration when comparing the SPDE and regTPS-KLE approaches is the relationship between model complexity (degrees of freedom) and computational efficiency. In the SPDE models the number of mesh nodes substantially exceeded the number of retained basis functions $M$ in the regTPS-KLE approach across all scenarios. Specifically, the regTPS-KLE method achieved dimension reductions ranging from 31\% to 58\% relative to the SPDE spatial degrees of freedom (Sce. 4 and Sce.1, respectively), while retaining 99\% of the cumulative KLE variance. This difference directly impacts computational performance: fewer degrees of freedom translate to more efficient MCMC sampling. The SPDE approach faces a fundamental trade-off between mesh resolution and computational tractability. Reducing the number of mesh nodes to improve computational speed often leads to convergence difficulties in MCMC sampling and potential loss of spatial resolution. Conversely, constructing high-quality, fine-resolution triangulations—necessary for capturing detailed spatial structure—produces accurate results but substantially increases computational burden. This trade-off between mesh resolution, computational efficiency, and MCMC convergence represents a key limitation of the SPDE approach when implemented via MCMC sampling \citep{cavieres2024approximated}. In contrast, the regTPS-KLE framework provides a more automated dimension reduction mechanism: the truncation parameter $M$ is selected objectively via variance retention, and the eigenvalue decay structure ensures that discarded high-frequency modes contribute negligible information. This adaptive truncation typically yields a parsimonious representation without requiring manual mesh tuning. Additionally, the two approaches differ in the number of hyperparameters requiring estimation. The SPDE model estimates three hyperparameters ($\sigma_u$, $\rho$, $\sigma_\varepsilon$), whereas the regTPS-KLE approach estimates two ($\alpha$, $\sigma_\varepsilon$). While this difference appears minor, the reduction from three to two hyperparameters marginally simplifies the posterior geometry and can improve MCMC mixing, particularly when using the non-centered parameterization described in Section~\ref{section3}. Moreover, the regularization parameter $\alpha$ in the regTPS-KLE framework has a direct interpretation as controlling the trade-off between data fidelity and smoothness, whereas the joint interpretation of $\sigma_u$ and $\rho$ in the Mat\'ern model is less intuitive for practitioners unfamiliar with spatial statistics.

\clearpage
\section{Conclusion}
\label{section7}

In the present work, we propose a novel approach to approximate a GRF using a TPS as covariance kernel. Since TPS is a conditional definite positive kernel, it cannot be used directly to approximate a GRF within the KLE method. Therefore, we establish a direct connection between the TPS to Hilbert-Schmidt kernels by considering the inverse of an elliptic operator $L_\alpha = \mathbf{I} + \alpha \Delta^{2}$, which yields  $L_\alpha^{-1}$, a compact, self-adjoint and positive definite operator. Under appropriate boundary conditions, the integral kernel associated with $L_\alpha^{-1}$ is a well- defined Hilbert-Schmidt kernel, and can therefore be directly used in the KLE method. This strategy offers several advantages, for example, it is not necessary consider specific prior distribution for hyperparameters in the kernel function (as the Mat\'ern or Exponential correlation function needs). This in turn, improves MCMC sampling by reducing issues related (possible) correlated parameters. From our numerical analysis, the regTPS-KLE accurately recovers the spatial fields simulated under the (assumed) Mat\'ern correlation function and performs well even when the fields are simulated from an Exponential correlation function. In the regTPS-KLE approach, the polynomial null space is given a unit prior variance, enabling the model to represent large-scale spatial trends without incurring a smoothing penalty. Explicitly handling the null space in this way provides a theoretical benefit compared to standard KLEs, which frequently omit the mean or handle it in a less consistent manner. Compared with the SPDE method, the regTPS-KLE approach reduces the problems of model convergence, since the SPDE approach often requires significant time to find correct priors for the hyperparameters and then achieve the stability in the chains. Furthermore, regTPS-KLE is computationally more efficient than SPDE models in most of the scenarios proposed. In the modeling of NO$_2$ concentrations, both approaches achieved accurate computation of posterior predictive distributions and spatial interpolation across the spatial domain (Germany). The results indicate similar large-scale spatial patterns in the posterior median of the predicted values, but with some differences in the upper tail of the posterior distribution (97.5\% quantile). The SPDE model predicts higher NO$_2$ concentrations in the northern and southeastern regions of Germany, whereas the regTPS-KLE model assigns higher values primarily to the western region. These differences arise from the distinct ways in which the two approaches approximate the GRF. In the SPDE framework, the marginal variance of the spatial field is governed by the hyperparameters $\rho$ and $\sigma_u$, which directly parameterize the Mat\'ern covariance structure and enter the construction of the sparse precision matrix $\mathbf{Q}$. In contrast, in the regTPS-KLE approach, the spatial field variance is determined by the regularization parameter $\alpha$ through the eigenvalue spectrum $\{\lambda_{k,\alpha}\}_{k=1}^M$, where each coefficient $z_k$ has prior variance $\lambda_{k,\alpha} = (1 + \alpha v_k)^{-1}$. This induces a smoothness-adaptive prior that allocates variance hierarchically: unit variance to polynomial null space modes ($k \leq d_{\text{null}}$) and progressively shrinking variance to higher-frequency components ($k > d_{\text{null}}$), with the decay rate controlled by $\alpha$ and the bending-energy eigenvalues $v_k$. Thus, while the SPDE approach uses a parametric Mat\'ern covariance family, the regTPS-KLE approach builds the covariance structure non-parametrically through the eigendecomposition of the regularized biharmonic operator. Despite their different methodological foundations, empirical results favor the regTPS-KLE model, which outperformed the SPDE model in its predictive accuracy, based on leave-one-out cross-validation, and computational efficiency in an MCMC algorithm.

\clearpage
\bibliographystyle{apalike}  
\bibliography{references}  

\begin{appendices}
\setcounter{equation}{0} 
\renewcommand{\theequation}{A\arabic{equation}} 

\section{Mass-Matrix Formulation and Generalized Ritz Eigenproblem}
\label{secA1}

In the main text, we employ the standard eigenvalue problem (Equation~\ref{eq33}) to obtain the discrete eigenfunctions of the penalty matrix $\mathbf{S}$. Here we provide the theoretical foundation for this choice by contrasting it with the fully consistent Galerkin formulation based on a generalized eigenvalue problem. The Galerkin discretization of the biharmonic operator presented in the main text relies on the bending-energy bilinear form
\begin{equation*}
a(u,v) = \int_{\mathcal D} (\Delta u)(\Delta v)\,d\mathbf s,    
\end{equation*}
which induces the stiffness matrix $\mathbf S$ defined in
Equation \eqref{eq32}. This formulation corresponds to measuring function roughness in the reproducing kernel Hilbert space (RKHS) associated with TPS. To obtain a fully consistent discrete analogue of the continuous eigenproblem $\Delta^2 \phi = v \phi$  in the $L^2(\mathcal D)$ sense, the Galerkin procedure must also incorporate the
$L^2(\mathcal D)$ inner product
\begin{equation*}
m(u,v) = \int_{\mathcal D} u(\mathbf s)\,v(\mathbf s)\,d\mathbf s,
\end{equation*}
whose discrete counterpart is the mass matrix
\begin{equation}\label{Aeq1}
\mathbf M_{jk} = \int_{\mathcal D}
\tilde\phi_j(\mathbf s)\,\tilde\phi_k(\mathbf s)\,d\mathbf s.
\end{equation}
Incorporating both bilinear forms leads to the generalized eigenvalue problem
\begin{equation}\label{Aeq2}
\mathbf S \boldsymbol{\psi}_k = v_k^{(N)} \, \mathbf M \boldsymbol{\psi}_k,
\end{equation}
where $(v_k^{(N)}, \boldsymbol{\psi}_k)$ are discrete Ritz pairs approximating the
eigenpairs $(v_k, \phi_k)$ of the continuous biharmonic operator $\Delta^2$. The associated discrete eigenfunctions are

\begin{equation}\label{Aeq3}
\hat\phi_k(\mathbf s) = \sum_{i=1}^{K} \psi_{ik} \,\tilde\phi_i(\mathbf s),
\end{equation}

and satisfy orthonormality with respect to the $L^2(\mathcal D)$ inner product,
\[
\int_{\mathcal D}
\hat\phi_k(\mathbf s)\,\hat\phi_\ell(\mathbf s)\,d\mathbf s
=
\boldsymbol{\psi}_k^{\top}\mathbf M\boldsymbol{\psi}_\ell
=
\delta_{k\ell}.
\]
Under standard regularity conditions, the eigenvalues $v_k^{(N)}$ and eigenfunctions
$\hat\phi_k$ converge to their continuous counterparts as the approximation space
$\mathcal V_N$ becomes dense in $H^2(\mathcal D)$. The Galerkin discretization of the regularized operator $L_\alpha = \mathbf I + \alpha \Delta^2$ in this setting yields the matrix representation
\begin{equation}\label{Aeq4}
\mathbf P_\alpha^{(M)} = \mathbf M + \alpha \mathbf S,
\end{equation}
where the superscript $(M)$ emphasizes the inclusion of the mass matrix. The corresponding discrete covariance operator is
\begin{equation}\label{Aeq5}
\mathbf K_\alpha^{(M)} = \left(\mathbf M + \alpha \mathbf S\right)^{-1}.
\end{equation}
Diagonalizing the generalized eigenproblem in Equation \eqref{Aeq2} yields
\begin{equation}\label{Aeq6}
\mathbf K_\alpha^{(M)} = \mathbf \Psi_M \boldsymbol{\Lambda}_\alpha \mathbf \Psi_M^{\top}, \qquad \lambda_{k,\alpha} =
\frac{1}{1 + \alpha v_k^{(N)}},
\end{equation}
where $\mathbf{\Psi}_M \in \mathbb{R}^{N \times N}$ has the eigenvectors $\boldsymbol{\psi}_k$ as its columns. Thus, $\mathbf K_\alpha^{(M)}$ constitutes a discrete KLE consistent with the $L^2(\mathcal D)$ geometry of the continuous problem. While the formal mass-matrix formulation in Equation\eqref{Aeq2} requires solving the generalized eigenvalue problem $\mathbf{S}\boldsymbol{\psi}_k = v_k^{(N)}\mathbf{M}\boldsymbol{\psi}_k$, our implementation adopts a computationally efficient approximation standard in TPS smoothing practice. Specifically, we compute eigenvalues $v_k$ from the standard eigenvalue problem

\begin{equation}\label{Aeq7}
\mathbf{S}\boldsymbol{\psi}_k = v_k\boldsymbol{\psi}_k,
\end{equation}

and employ these eigenvalues directly in the regularized covariance formula from Equation \eqref{Aeq6}:

\begin{equation}\label{Aeq8}
\lambda_{k,\alpha} = \frac{1}{1 + \alpha v_k}.
\end{equation}

This approach is justified as a computational approximation commonly adopted in practical TPS implementations and low-rank Gaussian process models \citep{wood2017generalized}. For irregular domains or highly clustered observation points, the mass matrix $\mathbf M$ is not diagonal, and the generalized eigenvalues reflect both roughness and local sampling density. By using the standard eigenvalue problem \eqref{Aeq7}, we rank modes purely by bending energy, which provides a stable, interpretable low-rank representation and allows the regularization parameter $\alpha$ to absorb any constant scaling. The null space of $\mathbf S$ (corresponding to polynomial trends with $v_k \approx 0$) is handled by assigning unit variance in the prior, $\lambda_k = 1$ for $k \leq M_P$, ensuring these components remain unpenalized as required by TPS theory \citep{duchon1977splines, green1987penalized, wood2017generalized}.

\section[alternative title goes here]{The SPDE method}\label{secA2}

Consider a GRF $u(\mathbf{s})$ with a Mat\'ern covariance function

\begin{equation}
\mathrm{Cov}(\mathbf{s}_{i}, \mathbf{s}_{j}) 
= \frac{\sigma^{2}_{u}}{2^{\nu-1}\Gamma(\nu)}
\left( \kappa \lVert \mathbf{s}_{i} - \mathbf{s}_{j} \rVert \right)^{\nu}
K_{\nu}\!\left( \kappa \lVert \mathbf{s}_{i} - \mathbf{s}_{j} \rVert \right),
\end{equation}

where $\Gamma(\cdot)$ is the Gamma function, $\kappa > 0$ is the spatial scale parameter, 
$\sigma_{u}$ is the marginal standard deviation, $\nu > 0$ controls smoothness, 
and $K_{\nu}$ is the modified Bessel function of the second kind. Following the results of \citet{whittle1954stationary,whittle1963stochastic}, the stationary solution of the stochastic partial differential equation (SPDE)

\begin{equation}
(\kappa^{2} - \Delta)^{\alpha/2}\big(\tau\, u(\mathbf{s})\big) 
= \mathcal{W}(\mathbf{s}), 
\qquad \mathbf{s} \in \mathbb{R}^{d},
\end{equation}

has a Mat\'ern covariance structure, where $\tau > 0$, $\alpha > d/2$, $\Delta$ is the Laplacian, and $\mathcal{W}$ is Gaussian spatial white noise. If the spatial domain $D \subset \mathbb{R}^{d}$ is bounded, then $u(\mathbf{s})$ can be approximated by a Gaussian Markov random field (GMRF) using the finite element method (FEM) \citep{lindgren2011explicit}. Specifically, the GRF is represented as

\begin{equation}
u(\mathbf{s}) = \sum_{k=1}^{n} \psi_{k}(\mathbf{s}) u_{k},
\end{equation}

where $\{\psi_{k}\}$ are piecewise linear FEM basis functions defined on a spatial triangulation of $D$, and $\mathbf{u} = (u_{1}, \ldots, u_{n})^{\top} \sim \mathcal{N}(\mathbf{0}, \mathbf{Q}^{-1})$. Here, $\mathbf{Q}$ is a sparse precision matrix constructed from FEM mass and stiffness matrices, denoted $\mathbf{C}$ and $\mathbf{G}$, respectively. The FEM matrices are defined as

\begin{align*}
C_{ii} &= \langle \psi_{i}, 1 \rangle, \\
G_{ij} &= \langle \nabla \psi_{i}, \nabla \psi_{j} \rangle, \\
\mathbf{K} &= \kappa^{2}\mathbf{C} + \mathbf{G}.
\end{align*}

Using the SPDE framework, the precision matrix $\mathbf{Q}$ can be constructed recursively as a function of $\kappa$ and $\alpha$:
\begin{align*}
\mathbf{Q}_{\alpha=1} &= \mathbf{K}, \\
\mathbf{Q}_{\alpha=2} &= \mathbf{K}\mathbf{C}^{-1}\mathbf{K}, \\
\mathbf{Q}_{\alpha \geq 3} &= \mathbf{K}\mathbf{C}^{-1}\mathbf{Q}_{\alpha-2}\mathbf{C}^{-1}\mathbf{K}.
\end{align*}

In particular, for $\alpha=2$, one obtains the explicit expression

\begin{equation}\label{Aeq12}
\mathbf{Q} = \tau^{2}\Big(\kappa^{4}\mathbf{C} 
+ 2\kappa^{2}\underbrace{\mathbf{G}}_{\mathbf{G}_{1}} 
+ \underbrace{\mathbf{G}\mathbf{C}^{-1}\mathbf{G}}_{\mathbf{G}_{2}}\Big),
\end{equation}

which is the standard formulation used in practice \citep{lindgren2015bayesian,bakka2018spatial}. 

\subsection{Parameterization}

The parameterization of the GRF commonly assumes a joint normal distribution for $\tau$ and $\kappa$ in log scale. Since $\kappa$ and $\tau$ have a joint influence on the variance marginal of the spatial random field, it is often more convenient to parameterize the model through the range $\rho$ and marginal standard deviation $\sigma_u$. The general relationships are:

\begin{equation}\label{Aeq13}
\rho = \frac{\sqrt{8\nu}}{\kappa} \hspace{6mm} \text{and} \hspace{6mm} \sigma_u = \sqrt{\frac{\Gamma(\nu)}{\tau^2 \kappa^{2\nu}(4\pi)^{d/2}\Gamma(\nu + d/2)}}.
\end{equation}

For the commonly used case with $d=2$ (two-dimensional spatial domain) and $\alpha=2$ (which corresponds to $\nu = \alpha - d/2 = 1$), these relationships simplify to:

\begin{equation}\label{Aeq14}
\kappa = \frac{\sqrt{8}}{\rho} \hspace{6mm} \text{and} \hspace{6mm} \tau = \frac{1}{\kappa \sigma_u} = \frac{\rho}{\sqrt{8}\sigma_u}.
\end{equation}

This simplified parameterization is used throughout our implementation.

\section{Model Formulations}\label{secA3}

We consider the following general model for spatial data:

\begin{equation}\label{Aeq15}
y_i = u(\mathbf{s}_i) + \varepsilon_i, \qquad \varepsilon_i \overset{\mathrm{iid}}{\sim} \mathcal{N}(0, \sigma_\varepsilon),
\end{equation}

where $y_i$ are the observed values at spatial locations $\mathbf{s}_i \in \mathbb{R}^2$, $u(\mathbf{s})$ is a GRF, and $\varepsilon_i \sim \mathcal{N}(0, \sigma_\varepsilon)$ (with $\sigma_\varepsilon > 0$). From this, we consider two approximations of $u(\mathbf{s})$: the SPDE method and the regTPS-KLE approach with the purpose to fit the model. 

\subsection{SPDE formulation}

\subsubsection{Non-centered parameterization}

To improve sampling efficiency and posterior geometry in MCMC sampling we employ a non-centered parameterization \citep{papaspiliopoulos2007general,betancourt2015hamiltonian}. Let $\mathbf{\tilde{u}} \sim \mathcal{N}(\mathbf{0}, \mathbf{Q}(\kappa)^{-1})$ be a spatial field. The centered field $\mathbf{u}$ is obtained through:

\begin{equation}\label{Aeq16}
\mathbf{u} = \frac{1}{\tau} \mathbf{\tilde{u}},
\end{equation}

which ensures that $\mathbf{u} \sim \mathcal{N}(\mathbf{0}, \tau^{-2}\mathbf{Q}(\kappa)^{-1})$ with the desired marginal standard deviation $\sigma_u = 1/(\tau\kappa)$. This parameterization decorrelates the spatial field from the hyperparameters $\kappa$ and $\tau$, substantially improving MCMC mixing and convergence. The spatial field at spatial locations is obtained via the projection matrix $\mathbf{A}_{\text{obs}} \in \mathbb{R}^{n \times m}$, which maps the $m$ mesh nodes to the $n$ spatial locations:

\begin{equation}\label{Aeq17}
u(\mathbf{s}_i) \approx (\mathbf{A}_{\text{obs}} \mathbf{u})_i, \quad i = 1,\ldots,n.
\end{equation}

which is the approximated GRF $u(\mathbf{s})$ at the spatial locations $\mathbf{s}_i$. 

\subsubsection{Priors}

We specify priors for the hyperparameters:

\begin{align*}
\sigma_\varepsilon &\sim \text{Cauchy}^+(0, s), \quad s = 5, \\
\rho^{-1} &\sim \text{Weibull}(\lambda_\rho^{-1}, 1), \\
\sigma_u &\sim \text{Exponential}(\lambda_{\sigma_u}),
\end{align*}

where $\text{Cauchy}^+(0, s)$ denotes a half-Cauchy distribution (the positive half of a Cauchy distribution with location 0 and scale $s$), and the Weibull distribution on $\rho^{-1}$ with shape parameter 1 corresponds to an exponential-like prior that can be parameterized via the probability statement $\Pr(\rho < \rho_0) = p_\rho$ through $\lambda_\rho = -\rho_0 \log(p_\rho)$. The exponential prior on $\sigma_u$ is specified via the probability statement $\Pr(\sigma_u > \sigma_{0,u}) = p_{\sigma_u}$ through $\lambda_{\sigma_u} = -\log(p_{\sigma_u})/\sigma_{0,u}$.

\subsubsection{Joint posterior distribution}

Combining the likelihood in Equation~\eqref{Aeq15} with the GMRF/SPDE prior and hyperpriors, the joint posterior distribution is:

\begin{equation}\label{Aeq18}
p(\mathbf{\tilde{u}}, \sigma_\varepsilon, \rho, \sigma_u \mid \mathbf{y}) 
\propto 
\left[ \prod_{i=1}^n \mathcal{N}\big(y_i \mid (\mathbf{A}_{\text{obs}}\mathbf{u})_i, \sigma^2_\varepsilon\big) \right]
\mathcal{N}\big(\mathbf{\tilde{u}} \mid \mathbf{0}, \mathbf{Q}(\kappa)^{-1}\big)
\pi(\sigma_\varepsilon)\pi(\rho)\pi(\sigma_u),
\end{equation}

where $\mathbf{u} = \mathbf{\tilde{u}}/\tau$ as in Equation~\eqref{Aeq16}, and $\kappa$ and $\tau$ are functions of $\rho$ and $\sigma_u$ through Equation~\eqref{Aeq13}.

\begin{figure}[!ht]
\centering
\includegraphics[width=10cm, height=10cm]{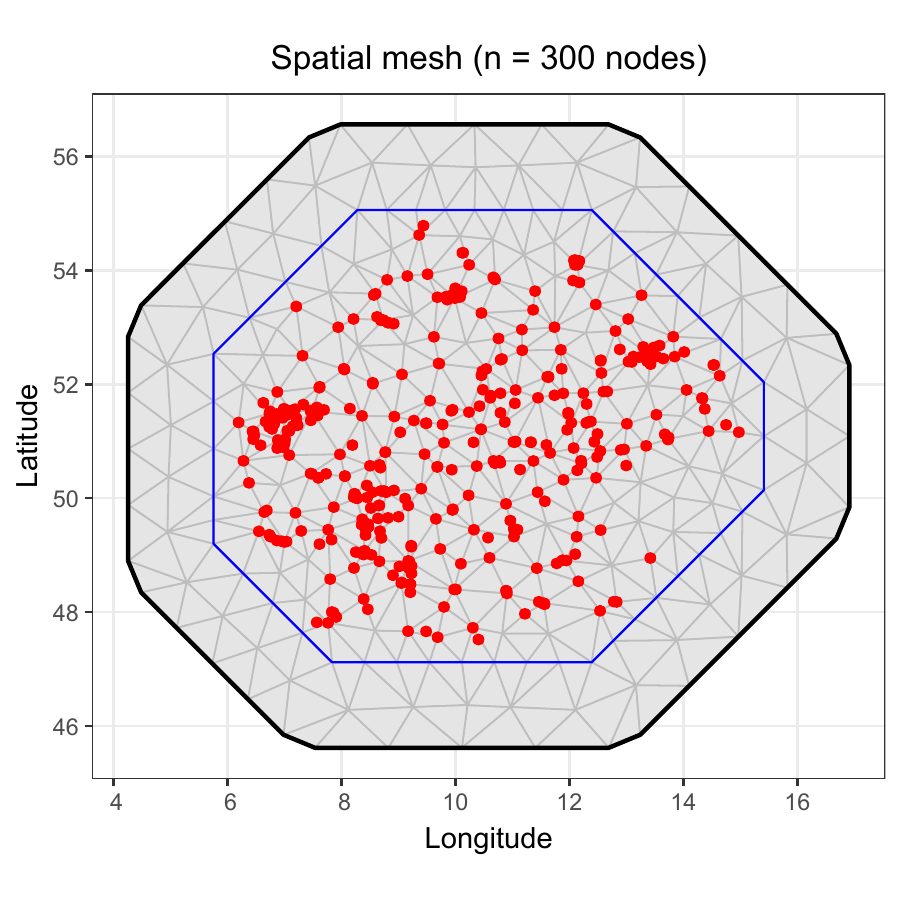}
\caption{Mesh constructed for the SPDE model in the real application of NO$_2$ modeling across Germany.}\label{fig:fig12}
\end{figure}

\subsection{regTPS--KLE formulation}

In the regTPS-KLE approach, the spatial field $u(\mathbf{s})$ is represented using the truncated KLE expansion:

\begin{equation}\label{Aeq19}
u(\mathbf{s}) = \sum_{k=1}^{M} z_k \, \hat{\phi}_k(\mathbf{s}),
\end{equation}

where $\{\hat{\phi}_k\}_{k=1}^M$ are the discrete eigenfunctions arising from the eigendecomposition of the TPS penalty matrix $\mathbf{S}$ (Equation~\ref{eq33}), and $z_k$ are the corresponding expansion coefficients. These eigenfunctions form an orthogonal basis in coefficient space, with the null space (first $M_{\text{null}} = d_{\text{null}}$ components) corresponding to low-order polynomials that are unpenalized by the TPS roughness penalty. We truncate the expansion at $M \leq K$ components to retain a specified proportion of total variance (typically 95--99\%, see Equation~\ref{eq39}).

\subsubsection{Non-centered parameterization}

To improve posterior geometry and computational efficiency, we adopt a non-centered parameterization. The transformation is:

\begin{equation}\label{Aeq20}
z_k =
\begin{cases}
\tilde{z}_k, & k \leq M_{\text{null}}, \\[6pt]
\displaystyle \frac{\tilde{z}_k}{\sqrt{1+\alpha v_k}}, & k > M_{\text{null}},
\end{cases}
\end{equation}

where $\tilde{z}_k \sim \mathcal{N}(0,1)$ are standard normal random variables, $v_k$ are the eigenvalues of the TPS penalty matrix $\mathbf{S}$, and $\alpha > 0$ is the regularization (smoothing) parameter. This parameterization corresponds to a Gaussian prior with precision matrix $\mathbf{P}_\alpha = \mathbf{I} + \alpha \mathbf{S}$ (Equation~\ref{eq42}), ensuring that null space components remain unpenalized ($\lambda_{k,\alpha} = 1$) while higher-frequency (rougher) modes are increasingly shrunk as $\alpha$ or $v_k$ increases ($\lambda_{k,\alpha} = (1 + \alpha v_k)^{-1}$).

Let $\boldsymbol{\Phi}_{\text{obs}} \in \mathbb{R}^{N \times M}$ denote the pre-computed KLE design matrix whose entries are $(\boldsymbol{\Phi}_{\text{obs}})_{ik} = \hat{\phi}_k(\mathbf{s}_i)$, i.e., the discrete TPS eigenfunctions evaluated at the observation locations (Equation~\ref{eq49}). The discretized spatial field at observation locations is:

\begin{equation}\label{Aeq21}
\mathbf{u} = \boldsymbol{\Phi}_{\text{obs}} \mathbf{z},
\end{equation}

where $\mathbf{z} = (z_1,\ldots,z_M)^\top$ with each $z_k$ defined through the transformation in Equation~\eqref{Aeq20}.

\subsubsection{Priors}

The full Bayesian specification is completed by assigning priors to the model parameters:

\begin{align*}
\sigma_\varepsilon &\sim \text{Exponential}(\lambda_\sigma),
\quad \lambda_\sigma = -\frac{\log(p_\sigma)}{\sigma_{0}}, \\
\log \alpha &\sim \mathcal{N}(\mu_\alpha, \sigma_\alpha^2), \\
\tilde{z}_k &\sim \mathcal{N}(0,1), \quad k = 1,\ldots,M. 
\end{align*}

The exponential prior on $\sigma_\varepsilon$ is a penalized complexity (PC) prior \citep{simpson2017penalising} with the probability statement $\Pr(\sigma_\varepsilon > \sigma_0) = p_\sigma$. We set $\sigma_0 = 0.5$ and $p_\sigma = 0.05$. The lognormal prior on $\alpha$ is specified through the normal prior on $\log \alpha$ with mean $\mu_\alpha = 0$ and standard deviation $\sigma_\alpha = 3$, allowing flexibility while maintaining weak informativeness and enabling the data to determine the appropriate degree of regularization.

\subsubsection{Hyperparameter Initialization}

To improve MCMC convergence, we initialize the regularization parameter $\alpha$ using a moment-matching approach. We estimate the signal variance as $\hat{\sigma}^2_{0} = \max(\text{Var}(y) - \sigma_\varepsilon^2, 0.1)$ and set the initial value:

\begin{equation}\label{Aeq23}
\alpha_{\text{init}} = \frac{\hat{\sigma}^2_{0}}{q_{0.25}(\{v_k : k > M_{\text{null}}, v_k > 10^{-10}\})},
\end{equation}

where $q_{0.25}$ denotes the 25th percentile of the non-zero penalty eigenvalues. This heuristic ensures that moderate-frequency components receive reasonable prior variance under the initial value, facilitating faster adaptation during sampling.

\subsubsection{Joint posterior distribution}

Given the observed values $y_i = u(\mathbf{s}_i) + \varepsilon_i$ with $\varepsilon_i \sim \mathcal{N}(0,\sigma_\varepsilon^2)$, the joint posterior distribution is:

\begin{equation}\label{Aeq24}
p(\tilde{\mathbf{z}}, \sigma_\varepsilon, \alpha \mid \mathbf{y})
\propto
\left[
\prod_{i=1}^N
\mathcal{N}\big(y_i \mid (\boldsymbol{\Phi}_{\text{obs}} \mathbf{z})_i, \sigma_\varepsilon^2\big)
\right]
\prod_{k=1}^{M}
\mathcal{N}\big(\tilde{z}_k \mid 0,1\big)
\,\pi(\sigma_\varepsilon)\,\pi(\alpha),
\end{equation}

where the likelihood depends on $\tilde{\mathbf{z}}$ through the transformation in Equation~\eqref{Aeq20}, and $\mathbf{z}$ is constructed by applying the non-centered transformation to each component.

\section{R Code for Data Simulation}\label{secA4}

To evaluate the SPDE and regTPS-KLE approaches under controlled conditions, we generate simulated spatial data considering a GRF with known Mat\'ern covariance. We consider four scenarios with sample sizes $N \in \{50, 100, 150, 200\}$, where observation locations $\mathbf{s}_i$ are uniformly distributed on the unit square $[0,1]^2$.

\begin{lstlisting}[language=R, caption={Simulation code for comparative study of SPDE and regTPS-KLE approaches}, label={lst:simulation}, basicstyle=\small\ttfamily, numbers=left, numberstyle=\tiny\color{gray}, keywordstyle=\color{blue}, commentstyle=\color{gray}\itshape, stringstyle=\color{red}, breaklines=true, frame=single]
#===================================================
#             Simulating the Data 
#===================================================
set.seed(1234)
base_N_sp <- 50
n_scenarios <- 4
dim_grid <- 30
sigma_u <- 1.0
sigma0_error <- 0.3
nu <- 1.5
rho <- 0.3

# Matern covariance function used for simulation
matern_cov <- function(coords, nu, rho, sigma2) {
  D <- as.matrix(dist(coords))
  D[D == 0] <- 1e-10
  scaling <- (sqrt(2 * nu) * D) / rho
  matern_part <- (2^(1 - nu)) / gamma(nu) * scaling^nu * besselK(scaling, nu)
  diag(matern_part) <- 1
  return(sigma2 * matern_part)
}

#====================================================
# Creating the lists for the SPDE/regTPS-KLE models
#====================================================
spde_models <- list()
regTPS_KLE_models <- list()

for (i in 1:n_scenarios) {
  N_sp <- base_N_sp * i
  
  # Create a common mesh and points for this scenario
  sp_points <- data.frame(s1 = runif(N_sp), s2 = runif(N_sp))
  sp_matrix <- as.matrix(sp_points)
  bound1 <- fmesher::fm_nonconvex_hull(sp_matrix)
  mesh <- fmesher::fm_rcdt_2d_inla(loc = sp_matrix, boundary = bound1, 
                                   refine = FALSE, plot.delay = NULL)
  
  # Create a common grid for this scenario
  grid_total <- expand.grid(s1 = seq(0, 1, length.out = dim_grid), 
                           s2 = seq(0, 1, length.out = dim_grid))
  
  # Simulate the TRUE latent field on the mesh nodes ONCE
  Cov_true <- matern_cov(mesh$loc, nu = nu, rho = rho, sigma2 = sigma_u^2)
  u_true <- as.numeric(MASS::mvrnorm(1, mu = rep(0, mesh$n), Sigma = Cov_true))
  
  # Create projection matrices to get the true field at other locations
  A_obs_proj <- inla.spde.make.A(mesh = mesh, loc = sp_matrix)
  A_grid_proj <- inla.spde.make.A(mesh = mesh, loc = as.matrix(grid_total))
  
  # Project the true field to observation points and grid points
  u_true_sp <- as.numeric(A_obs_proj %*% u_true)
  u_true_grid <- as.numeric(A_grid_proj %*% u_true)
  
  # Add noise to the observations
  y_obs <- u_true_sp + rnorm(N_sp, 0, sigma0_error)
  
  # Run SPDE model first
  obj_spde <- run_tmb_spde(N_sp, dim_grid, sp_points, mesh, y_obs, 
                          u_true_sp, u_true_grid, Cov_true)
  
  # Get number of mesh nodes from SPDE
  n_mesh_nodes <- mesh$n
  
  #====================================
  # COMPARISON STRATEGY FOR regTPS-KLE
  #====================================
  # Maximum k_basis for TPS (must be < N_sp)
  k_basis_max <- floor(0.99 * N_sp)   # 99% of data points
  k_basis_max <- max(k_basis_max, 10) # At least 10 basis functions
  
  # Run TPS model
  obj_tps <- run_tmb_tps(N_sp, dim_grid, sp_points, mesh, y_obs, 
                        u_true_sp, u_true_grid, 
                        k_basis = k_basis_max, 
                        Cov_true, 
                        variance_threshold = 0.99)
  
  if(obj_tps$M_truncation < n_mesh_nodes) {
    reduction_pct <- round((1 - obj_tps$M_truncation / n_mesh_nodes) * 100, 1)
  }
  
  spde_models[[i]] <- obj_spde
  regTPS_KLE_models[[i]] <- obj_tps
}
\end{lstlisting}

\end{appendices}

\end{document}